%% file: long.tex
\documentclass[twocolumn,twoside]{IEEEtran}

\usepackage{times}
\usepackage{amsmath}  
\usepackage{amssymb}  
\usepackage{theorem}  
\usepackage{cite}     
\usepackage{upref}
\usepackage{amsfonts}
\usepackage{verbatim}
\usepackage[dvipsnames,usenames]{color}
\usepackage{cite}
\usepackage[inline]{enumitem}
\usepackage[format=plain,
            font=it]{caption}
\usepackage{balance}
\usepackage{nicefrac}

\usepackage{algorithm}
\usepackage{times,color}
\usepackage{enumitem}
\usepackage{float}
\usepackage{color}
\usepackage{mathtools}
\usepackage{graphicx}
\usepackage{listings}
\usepackage{tikz}
\usepackage{multicol,multirow}
\usepackage{psfrag}
\usepackage{bbm}
\usepackage{verbatim}
\usepackage{soul}

\usepackage{pgfplots}
\usepackage{hhline}
\usepackage{xcolor,colortbl}
\usepackage{algorithm}
\usepackage{varwidth}
\usepackage[noend]{algpseudocode}
\usepackage{comment}

\usepackage{enumitem}
\usepackage{pgf}

\usetikzlibrary{%
	arrows,%
	automata,%
	calc,%
	patterns,%
	plotmarks,%
	positioning,%
	shapes.geometric,%
	shapes,%
	snakes,%
	decorations.pathmorphing,%
	decorations.shapes,%
	graphs,%
	matrix,%
	fit,%
	chains,%
	arrows.meta, %
}

\algnewcommand{\Initialize}[1]{%
	\State \textbf{Initialize:}
	\Statex \hspace*{\algorithmicindent}\parbox[t]{.8\linewidth}{\raggedright #1}
}

\input{commands_loglog}

\begin{document}

\title{\textbf{Multiple Criss-Cross Insertion and Deletion Correcting Codes}}%
\author{%
	\IEEEauthorblockN{{\textbf{Lorenz Welter}},         
	    {\textbf{Rawad Bitar}},
                {\textbf{Antonia Wachter-Zeh}}, 
		and {\textbf{Eitan Yaakobi}}
		} \\
	\thanks{LW, RB and AW-Z are with the Institute for Communications Engineering, Technical University of Munich (TUM), Germany. Emails: \{rawad.bitar, lorenz.welter, antonia.wachter-zeh\}@tum.de.}
	\thanks{EY is with the CS department of Technion --- Israel Institute of Technology, Israel. Email: yaakobi@cs.technion.ac.il.}
	
	\thanks{This project has received funding from the European Research Council (ERC) under the European Union’s Horizon 2020 research and innovation programme (grant agreement No. 801434) and from the Technical University of Munich - Institute for Advanced Studies, funded by the German Excellence Initiative
and European Union Seventh Framework Programme under Grant Agreement
No. 291763.}\vspace{-4ex}
\thanks{Preliminary results of this work is published in ISIT 2021~\cite{ISITcriss}.}}

\maketitle

\begin{abstract}

This paper investigates the problem of correcting multiple criss-cross insertions and deletions in arrays. More precisely, we study the unique recovery of $n \times n$ arrays affected by \emph{$\etot$-criss-cross deletions} defined as any combination of $\erow$ row and $\ecol$ column deletions such that $\erow + \ecol = \etot$ for a given $t$. We show an equivalence between correcting $\etot$-criss-cross deletions and $\etot$-criss-cross insertions and show that a code correcting $\etot$-criss-cross insertions/deletions has redundancy at least $\etot n + \etot \log n - \log(\etot!)$. Then, we present an existential construction of a $\etot$-criss-cross insertion/deletion correcting code with redundancy bounded from above by $\etot n + \mathcal{O}(\etot^2 \log^2 n)$. The main ingredients of the presented code construction are systematic binary $\etot$-deletion correcting codes and Gabidulin codes. The first ingredient helps locating the indices of the inserted/deleted rows and columns, thus transforming the insertion/deletion-correction problem into a row/column erasure-correction problem which is then solved using the second ingredient.

\end{abstract}

\section{Introduction} \label{sec:intro}

\input{intro}

\section{Definitions and Preliminaries}\label{sec:def}
\input{definitions.tex}

\section{Equivalence between Insertion and Deletion Correction}\label{sec:equiv}
\input{equivalence}

\section{Upper Bound on the cardinality}\label{sec:bounds}
\input{card-bound_loglog.tex}

 \section{Code Construction}\label{sec:cons}
\input{construction_loglog}

 \section{Decoder}\label{sec:decoder}

\input{decoder_loglog}

 \section{Redundancy}\label{sec:red}
\input{redundancy_loglog}

\section{Conclusion}
\input{conclusion_loglog}
\bibliographystyle{IEEEtran}
\bibliography{refs}

\balance
\end{document}

%% file: commands_loglog.tex
\newcommand{\ie}{{i.e., }}

\newcommand{\be}[1]{\begin{equation}\label{#1}}
\newcommand{\ee}{\end{equation}}

\newcommand{\bc}{\begin{center}}
\newcommand{\ec}{\end{center}}

\newcommand{\cC}{{\cal C}}

\newcommand{\bfk}{{\boldsymbol k}}

\newcommand{\bfr}{{\boldsymbol r}}

\newcommand{\bfB}{{\mathbf B}}
\newcommand{\bfC}{{\mathbf C}}
\newcommand{\bfD}{{\mathbf D}}
\newcommand{\bfE}{{\mathbf E}}
\newcommand{\bfF}{{\mathbf F}}
\newcommand{\bfG}{{\mathbf G}}

\newcommand{\bfI}{{\mathbf I}}

\newcommand{\bfL}{{\mathbf L}}

\newcommand{\bfR}{{\mathbf R}}

\newcommand{\bfT}{{\mathbf T}}

\newcommand{\bfW}{{\mathbf W}}
\newcommand{\bfX}{{\mathbf X}}
\newcommand{\bfY}{{\mathbf Y}}
\newcommand{\bfZ}{{\mathbf Z}}

\newcommand{\xij}{{X_{i,j}}}

\renewcommand{\leq}{\leqslant}

\renewcommand{\geq}{\geqslant}

\newcommand{\N}{\mathbb{N}}

\newcommand{\Cref}[1]{Co\-rol\-la\-ry\,\ref{#1}}

\newtheorem{theorem}{Theorem} 
\newtheorem{lemma}[theorem]{Lemma} 
\newtheorem{lem}[theorem]{Lemma}
\newtheorem{definition}{Definition} 
 
\newtheorem{cor}[theorem]{Corollary}

\newtheorem{claim}[theorem]{Claim}

\newtheorem{construction}{Construction}

\definecolor{Codecolor}{named}{White}

\newcommand{\Copen}{\mbox{\{\kern-5.50pt\{}}
\newcommand{\Cclose}{\mbox{\}\kern-5.50pt\}}}
\newcommand{\Cslash}{\mbox{$\backslash\kern-6.02pt\backslash$}}

\newcommand{\oi}{\smash{\Sigma}}

\newcommand{\sigmatn}{\smash{\Sigma ^{n \times n} }}

\newcommand{\dball}{\ensuremath{\mathbb{D}}}

\newcommand{\iball}{\ensuremath{\mathbb{I}}}

\newcommand{\tr}{\ensuremath{t_{\mathrm{r}}}}
\newcommand{\tc}{\ensuremath{t_{\mathrm{c}}}}

\newcommand{\etot}{\ensuremath{t}}
\newcommand{\erow}{\ensuremath{\etot_{\mathrm{r}}}}
\newcommand{\ecol}{\ensuremath{\etot_{\mathrm{c}}}}
\newcommand{\s}{\ensuremath{s}}
\renewcommand{\a}{\ensuremath{a}}
\newcommand{\ro}{\ensuremath{\rho}}
\renewcommand{\k}{\ensuremath{\kappa}}

\newcommand{\wcons}{window constraint}

\newcommand{\redopt}{\ensuremath{R_\mathrm{B}(n,\etot)}}

\newcommand{\idmat}[1]{\ensuremath{\bfI _{#1}}}
\newcommand{\allone}[1]{\ensuremath{\mathbf{1}_{#1}}}

\newcommand{\econstarr}[1]{\ensuremath{\mathbf{M}^{(#1)}}}
\newcommand{\tileconstarr}[1]{\ensuremath{\widetilde{\mathbf{M}}^{(#1)}}}

\newcommand{\allzero}[1]{\ensuremath{\mathbf{0}_{#1}}}
\newcommand{\leftmarg}[1]{\ensuremath{\bfL}_{#1}}
\newcommand{\topmarg}[1]{\ensuremath{\bfT}_{#1}}
\newcommand{\leftm}[1]{\ensuremath{\bfL}^{#1}}

\newcommand{\winarr}[1]{\ensuremath{\mathcal{W}_\etot(#1)}}

\newcommand{\gabi}[2]{\ensuremath{\mathcal{C}_{\mathrm{Gab}}(#1,#2)}}
\newcommand{\leftmargo}{\ensuremath{\bfL}}
\newcommand{\topmargo}{\ensuremath{\bfT}}

\newcommand{\edelarr}[1]{\ensuremath{\mathcal{D} ^{(#1)} _{\etot}}}

\newcommand{\locaset}{\ensuremath{\mathcal{L}_\etot (n)}}

\newcommand{\codedef}{\ensuremath{\mathcal{C}_{\etot,n}}}

\newcommand{\vertedel}[1]{\ensuremath{\mathcal{V}_{\etot}(#1)}}
\newcommand{\horzedel}[1]{\ensuremath{\mathcal{H}_{\etot}(#1)}}
\newcommand{\margarr}[1]{\ensuremath{\mathcal{E}_{\etot}(#1)}}
\newcommand{\markerset}[1]{\ensuremath{\mathcal{M}_{\etot}(#1)}}

\newcommand{\rdset}{\ensuremath{\mathcal{I}^{(\erow)}}}
\newcommand{\cdset}{\ensuremath{\mathcal{I}^{(\ecol)}}}

\newcommand{\twot}{\ensuremath{k}}

\newcommand{\red}{\ensuremath{R(n,\etot)}}
\newcommand{\redgab}{\ensuremath{R_{\mathrm{G}}(n,\etot)}}
\newcommand{\redloca}{\ensuremath{R_\mathcal{L}(n,\etot)}}

\newcommand{\redho}{\ensuremath{R_\mathcal{H}(n,\etot)}}
\newcommand{\redve}{\ensuremath{R_\mathcal{V}(n,\etot)}}
\newcommand{\redw}[1]{\ensuremath{R_\mathcal{W}(\ell,#1)}}
\newcommand{\redmanab}{\ensuremath{R_\mathcal{E}(n,\etot)}}
\newcommand{\redmarker}{\ensuremath{R_\mathcal{M}(n,\etot)}}

\definecolor{lightseagreen}{rgb}{0.13, 0.7, 0.67}
\definecolor{darkolivegreen}{rgb}{0.33, 0.42, 0.18}
\definecolor{mediumpersianblue}{rgb}{0.0, 0.4, 0.65}
\definecolor{bostonuniversityred}{rgb}{0.8, 0.0, 0.0}
\definecolor{brilliantrose}{rgb}{1.0, 0.33, 0.64}
\definecolor{dartmouthgreen}{rgb}{0.05, 0.5, 0.06}
\definecolor{debianred}{rgb}{0.84, 0.04, 0.33}
\definecolor{amber}{rgb}{1.0, 0.49, 0.0}
\definecolor{cadmiumyellow}{rgb}{1.0, 0.96, 0.0}
\definecolor{cocoabrown}{rgb}{0.82, 0.41, 0.12} 

\pgfdeclarepatternformonly{south west lines}{\pgfqpoint{-0pt}{-0pt}}{\pgfqpoint{3pt}{3pt}}{\pgfqpoint{3pt}{3pt}}{
        \pgfsetlinewidth{0.4pt}
        \pgfpathmoveto{\pgfqpoint{0pt}{0pt}}
        \pgfpathlineto{\pgfqpoint{3pt}{3pt}}
        \pgfpathmoveto{\pgfqpoint{2.8pt}{-.2pt}}
        \pgfpathlineto{\pgfqpoint{3.2pt}{.2pt}}
        \pgfpathmoveto{\pgfqpoint{-.2pt}{2.8pt}}
        \pgfpathlineto{\pgfqpoint{.2pt}{3.2pt}}
        \pgfusepath{stroke}}

\newcommand{\review}[2]{{\textcolor{black}{#1}}}

\newcommand{\LastCheck}[1]{{\color{black}{#1}}}

\newcommand{\insdel}{indel}
\newcommand{\Insdel}{Indel}

\newcommand{\tcc}{\etot-criss-cross \review{\insdel}{} code}
\newcommand{\tccs}{\etot-criss-cross \review{\insdel}{} codes}

\def \th{^\text{th}}

%% file: intro.tex
Codes correcting insertions and deletions have recently witnessed an increased attention due to their application in DNA-based storage systems, file synchronization, and communication systems~\cite{Heckel_A-Characterization-of-the-DNA-Storage-Channel_19,SalaSchoeny-Sync_TransComm,venkataramanan2010interactive,7185447,yazdi2013deterministic,ma2011efficient,DolecekAnan_Sync_2007}. The problem of correcting insertions and deletions, referred to as \emph{\insdel}\ errors, dates back to the 1960s. In~\cite{Levenshtein-binarycodesCorrectingDeletions}, Levenshtein defined the notion of $\etot$-deletion-correcting codes and showed that a code can correct any combination of $\etot$ \insdel s if and only if it can correct $\etot$ deletions. The main property of the codes that is usually optimized is the redundancy defined as $R\triangleq n-\log|\cC|$ where $n$ is the length of the codewords in $\cC$ and $|\cC|$ is the cardinality of the code. Levenshtein bounded the redundancy of any binary $\etot$-\insdel-correcting code  from below by $\etot \log n - \mathcal{O}(1)$. Moreover, he proved that the Varshamov-Tenengolts codes \cite{VarshTene-SingleDeletion1965}, originally designed to correct a single asymmetric error, can also correct a single \insdel\ and have redundancy roughly $\log(n+1)$ bits. 

Several recent works studied the problem of constructing binary $\etot$-\insdel-correcting codes, for $t>1$, with redundancy approaching Levenshtein's bound~\cite{GuruswamiWang-HighNoiseHighRateDeletions_2017,brakensiek2017efficient,hanna2018guess,Gabrys-TwoDeletions_2018,Sima-TwoDeletions_2019,SimaBruck-kDeletions_2020,guruswami2020explicit}. Of particular importance to us is the work of Sima \emph{et al.}~\cite{sima2020systematictdel} in which the authors present a binary systematic $\etot$-\insdel-correcting code with redundancy $4 \etot \log (n) + o \log(n)$. This code can correct any combination of $\etot$ \insdel\ errors.

This paper considers the problem of coding for \insdel s in the two-dimensional space. The motivation stems from the two-dimensional erasure and substitution correction problem where it has been shown that leveraging the structure of the array is more beneficial than applying one-dimensional error correcting codes on each dimension of the array. The deletion (and clearly also the {\insdel}) correction problem is however more involved due to the loss of synchronization in the locations of the inserted and deleted rows and columns. Along this line of thought, the trace-reconstruction problem, which is related to coding for deletions, is investigated for the two-dimensional space in~\cite{krishnamurthy2019trace}. Moreover, coding for deletions over the two-dimensional space is also considered in~\cite{SchoenyWachterzehGabrysYaakobi-BurstDeletions-journal,smith2017interleaved,bakirtas2021database}. In~\cite{SchoenyWachterzehGabrysYaakobi-BurstDeletions-journal,smith2017interleaved} codes that can correct bursts of deletions in the one-dimensional space are constructed. The main idea is to view the codeword as a binary array and use the structure of that array to detect and correct bursts of deletions that happen in the one-dimensional codeword. In~\cite{bakirtas2021database}, the authors consider the problem of database matching under column deletions. 

Given a certain number of deletions $\etot$ and an array $\bfX$, we assume that the array can be affected by any combination of $\erow$ row and $\ecol$ column deletions such that $\erow+\ecol = \etot$. This type of deletions are referred to as \emph{$\etot$-criss-cross deletions}. We define \emph{$\etot$-criss-cross insertions} similarly. Our goal is to construct codes that can uniquely recover the array $\bfX$ from any $\etot$-criss-cross deletion or any $\etot$-criss-cross insertion and we refer to these codes as \emph{\tccs}. We borrow this terminology from previous works that studied the problem of correcting criss-cross erasures and substitution errors in the two-dimensional space, e.g., \cite{roth1991maximum, gabidulin2008crisscrosserasure, LundGabidulinHonary-NewFamilyOptimalCorrectingTermRankErrors_2000,Sidorenko-ClassCorrectingErrorsLatticeConfiguration_1976,BlaumBruck-ArrayCodesCorrectionCrisscrossErrors_IEEE-IT2000,Gabidulin-OptimalArrayCorrectingCodes_1985,Roth-ProbabilisticCrisscrossErrorCorrection_1997,wachterzeh2017listdecodingcrisscross}.

The first works to study the criss-cross deletion problem were \cite{bitar2020criss,ISITAcriss,manabu2020delarrays}. In \cite{bitar2020criss,ISITAcriss}, we investigated the problem of correcting exactly one row and one column insertion/deletion in arrays. We showed that the redundancy of codes designed for this special case is bounded from below by $2n + 2 \log n - \mathcal{O}(1)$. We also presented an existential and an explicit construction with redundancy approximately $2 \log n$ and $7 \log n$ far from the lower bound, respectively. Furthermore, we showed that, for $\tr=\tc$, a code can correct any $\etot$-criss-cross deletion if and only if it can correct any $\etot$-criss-cross insertion. In~\cite{manabu2020delarrays}, Hagiwara constructed codes correcting criss-cross deletions with at most $\erow$ row deletions and at most $\ecol$ column deletions, for given values of $\erow$ and $\ecol$. The constructed codes have redundancy in the order of $n (\erow^2+\ecol^2+(\erow+\ecol) \log n)$. The construction splits the array into locators and information part. The locators are carefully structured arrays that can exactly recover the index of any deleted rows and columns in the array. Then, a tensor-product erasure-correcting code is used to recover the lost symbols in the information part.

Our contributions in this paper can be summarized as follows. We present an asymptotic upper bound (in the code length) on the cardinality of \tccs. Our bound implies that the redundancy of any \tcc\ is bounded from below by approximately $\etot n + \etot \log n$. We extend the equivalence between correcting deletions and insertions to the general $\etot$-criss-cross deletion model considered in this paper. Then, we construct existential \tccs\ based on locator arrays, binary systematic $\etot$-deletion correcting codes, and Gabidulin codes. We also show that this code can correct $\etot$-criss-cross insertions by providing an explicit decoder. The main improvements of our construction over the one in~\cite{manabu2020delarrays} is to use a collection of binary deletion-correcting codes to locate the indices of the deleted columns and a Gabidulin code to correct the erasures. This significantly reduces the redundancy of the code. However, small locator arrays are still needed to complement the deletion-correcting codes. Then, the deletion-correction problem is transformed into a row/column erasure-correction problem which can be solved by using Gabidulin codes that have optimal redundancy for row/column erasure-correction~\cite{gabidulin2008crisscrosserasure}. The redundancy of the presented construction is $\etot n + \mathcal{O}(\etot^2 \log^2 n)$. For the considered problem setting, we substantially improve upon the current state-of-the-art construction of~\cite{manabu2020delarrays} that needs a redundancy of approximately $2n \cdot (\etot^2+\etot \log n )$ in this setting.

%% file: definitions.tex
This section formally defines the codes and notations that are used throughout this paper. Let $\Sigma \triangleq \{0,1\}$ be the binary alphabet. We denote by $\sigmatn$ the set of all binary arrays of dimension $n \times n$. 
All logarithms are base $2$ unless otherwise indicated.

For an integer $n \in \N$, the set $\{ 1,\ldots , n \}$ is denoted by $[n]$. For an array $\bfX \in \sigmatn$ and $i,j \in [n]$, we refer to the entry of $\bfX$ positioned at the $i\th$ row and the $j\th$ column by $\xij$. We denote the $i\th$ row and the $j\th$ column of $\bfX$ by $\bfX_{i,[n]}$ and $\bfX_{[n],j}$, respectively. Similarly, we denote by $\bfX_{[i_1:i_2],[j_1:j_2]}$ the subarray of $\bfX$ formed by rows $i_1$ to $i_2$ and their corresponding entries from columns $j_1$ to $j_2$. \LastCheck{We denote by $\bfX^T$ the transpose of the array $\bfX$.} Moreover, for two arrays $\bfX \in \Sigma ^{n \times m_1}$ and $\bfY \in \Sigma ^{n \times m_2}$ we denote by $\bfZ = (\bfX \mid \bfY)$ the concatenation of these two arrays with $\bfZ \in \Sigma ^{n \times (m_1 + m_2)}$. For any binary array $\bfX$, we refer to the complement of $\bfX$, \ie every bit in $\bfX$ is flipped, by $\overline{\bfX}$. In an array $\bfX \in \Sigma^{n\times m}$, a column-run of length $r$ is defined as a sequence of $r$ consecutive equal columns $\bfX_{[n],j}=\bfX_{[n],j+1}=\dots = \bfX_{[n],j+r-1}$. Row-runs in an array $\bfX$ are defined similarly. Given a vector $\mathbf{x}\in \Sigma^n$ a run of length $r$ in $\mathbf{x}$ is defined as a sequence of $r$ consecutive equal bits $x_{i}=x_{i+1}=\dots=x_{i+r-1}$.

For positive integers $\tr, \tc$ we define a $(\tr,\tc)$-criss-cross deletion in a binary array $\bfX$ to be the deletion of any $\tr$ rows and $\tc$ columns of $\bfX$. For a positive integer $\etot$, we define a \emph{$\etot$-criss-cross deletion} in a binary array $\bfX$ to be the collection of all $(\tr,\tc)$-criss-cross deletions in $\bfX$ such that $\erow + \ecol = \etot$. Further, $(\tr,\tc)$-criss-cross insertion and a \emph{$\etot$-criss-cross insertion} are defined similarly. We denote by $\dball_{\erow,\ecol}(\bfX)$ the set of all arrays that result from $\bfX$ after a $(\erow,\ecol)$-criss-cross deletion (i.e., the two-dimensional deletion ball\footnote{{Strictly speaking, the set $\dball_{\erow,\ecol}(\bfX)$ must be called the two-dimensional deletion \emph{sphere} of $\bfX$. However, we abuse terminology and refer to this set as the deletion ball to follow the nomenclature used by the literature on deletion-correcting codes. The same holds for the set $\iball_{\erow,\ecol}(\bfX)$.}}).
In a similar way we define the set $\iball_{\tr,\tc}(\bfX)$ for the insertion case. We refer to $\widetilde{\bfX}$ as the array resulting from a $\etot$-criss-cross deletion or insertion in $\bfX$, where the number and type of errors (deletions or insertions) that happened in $\bfX$ is clear from the context.
A code $\cC\subseteq\sigmatn$ that can correct any $(\tr,\tc)$-criss-cross deletion or any $(\tr,\tc)$-criss-cross insertion is called a \emph{$(\tr,\tc)$-criss-cross \insdel-correcting code}. A {\emph{$\etot$-criss-cross \insdel-correcting code}} is defined similarly. We abbreviate those codes as $(\tr,\tc)$-criss-cross \insdel\ code and \emph{\tcc}, respectively.
Throughout this paper we assume that $\etot$ is a constant with respect to~$n$.
We write $f(n) \approx g(n)$,  $f(n) \lesssim g(n)$, and $f(n) \gtrsim g(n)$ if the equality or inequality holds for $n \to \infty$.

%% file: equivalence.tex
In the following, we show the equivalence between $t$-criss-cross deletion-correcting codes and $t$-criss-cross insertion-correcting codes (Theorem~\ref{theorem:equiv}). The proof of Theorem~\ref{theorem:equiv} follows by first showing that the equivalence holds for all $(\erow,\erow+c)$-criss-cross \insdel\ codes, where $c$ is a positive integer. Then, by symmetry the equivalence holds for $(\ecol+c,\ecol)$-criss-cross \insdel\ codes which completes the proof. 

\begin{theorem} \label{theorem:equiv}
A code $\cC\subset \sigmatn$ is a $t$-criss-cross deletion-correcting code if and only if $\cC$ is a $t$-criss-cross insertion-correcting code.
\end{theorem}

We need the following results from~\cite{bitar2020criss} showing that any $(\erow,\erow)$-criss-cross deletion-correcting code can also correct insertions and extending the properties of balls intersections from the one-dimensional space to the two-dimensional space for only one \insdel.

\begin{theorem}[\hspace{-0.05ex}\cite{bitar2020criss}]\label{thm:tinsdel}
 For all integers $\erow \in [n-1]$, a code $\cC\subset\oi ^{n\times n}$ is a $(\erow,\erow)$-criss-cross deletion-correcting code if and only if it is a $(\erow,\erow)$-criss-cross insertion-correcting code.
\end{theorem}

 \begin{lemma}[\hspace{-0.05ex}\cite{bitar2020criss}] \label{lemma:equivsimpl}
For a positive integer $m$ and two arrays $\bfX\in \oi^{m\times m}$ and $\bfY\in \oi^{m\times m}$, 
\begin{align*}
    &\dball_{1,0}(\bfX)\cap\dball_{1,0}(\bfY)=\emptyset \Leftrightarrow \iball_{1,0}(\bfX)\cap\iball_{1,0}(\bfY)=\emptyset, \\
    &\dball_{0,1}(\bfX)\cap\dball_{0,1}(\bfY)=\emptyset \Leftrightarrow \iball_{0,1}(\bfX)\cap\iball_{0,1}(\bfY)=\emptyset.
\end{align*}
 \end{lemma}
 
\begin{lemma}[\hspace{-0.05ex}\cite{bitar2020criss}] \label{lemma:equiv}
For a positive integer $m$ and two arrays $\bfX\in \oi^{(m+1)\times m}$ and $\bfY\in \oi^{m\times (m+1)}$, 
\begin{align*}
    &\dball_{1,0}(\bfX)\cap\dball_{0,1}(\bfY)=\emptyset \Leftrightarrow \iball_{0,1}(\bfX)\cap\iball_{1,0}(\bfY)=\emptyset.
\end{align*}
 \end{lemma}
 
We now use the previous results to prove Theorem~\ref{theorem:equiv}.

\begin{IEEEproof}[Proof of Theorem~\ref{theorem:equiv}]
The goal is to show that for any $\bfX_1,\bfX_2\in \sigmatn$ and any two integers $\erow$ and $\ecol$ such that $\erow+\ecol=\etot$, the following holds.
\begin{align*}
    \dball_{\erow,\ecol}(\bfX_1)\cap \dball_{\erow,\ecol}(\bfX_2) = \emptyset \Leftrightarrow \iball_{\erow,\ecol}(\bfX_1)\cap \iball_{\erow,\ecol}(\bfX_2) = \emptyset.
\end{align*}

We only show the equivalence for $(\erow,\erow+c)$-criss-cross \insdel\ codes, i.e., for any $\bfX_1,\bfX_2\in \sigmatn$ and any integers $\erow$ and $c$ such that $2\erow+c = t$ we have 
\begin{align*}
    &\dball_{\erow,\erow+c}(\bfX_1)\cap \dball_{\erow,\erow+c}(\bfX_2)=\emptyset \\
    &\Leftrightarrow \iball_{\erow,\erow+c}(\bfX_1)\cap \iball_{\erow,\erow+c}(\bfX_2)=\emptyset .
\end{align*}

Assuming that the equivalence holds for $(\erow,\erow+c)$-criss-cross \insdel\ codes, then the following implies that the equivalence holds for $(\erow+c,\erow)$-criss-cross \insdel\ codes which completes the proof.
\begin{align*}
    &\dball_{\erow,\erow+c}(\bfX_1)\cap \dball_{\erow,\erow+c}(\bfX_2) = \emptyset \\
    &\stackrel{(a)}{\Leftrightarrow} \dball_{\erow+c,\erow}(\bfX_1^T)\cap \dball_{\erow+c,\erow}(\bfX_2^T) = \emptyset\\
&\stackrel{(b)}{\Leftrightarrow} \iball_{\erow+c,\erow}(\bfX_1^T)\cap \iball_{\erow+c,\erow}(\bfX_2^T) = \emptyset\\
&\stackrel{(c)}{\Leftrightarrow} \iball_{\erow,\erow+c}(\bfX_1)\cap \iball_{\erow,\erow+c}(\bfX_2) = \emptyset.
\end{align*}
The statements (a) and (c) follow trivially by examining the transpose of the arrays $\bfX_1$ and $\bfX_2$. The statement (b) follows from the assumed equivalence.

The proof of equivalence for $(\erow,\erow+c)$-criss-cross \insdel\ codes proceeds by contraposition, i.e., we show that $\dball_{\erow,\erow+c}(\bfX_1)\cap \dball_{\erow,\erow+c}(\bfX_2) \neq \emptyset$ if and only if $\iball_{\erow,\erow+c}(\bfX_1)\cap \iball_{\erow,\erow+c}(\bfX_2) \neq \emptyset$. In what follows, we only show the ``only if" parts since the ``if" parts follow similarly.

We prove the equivalence for $(\erow,\erow+c)$-criss-cross \insdel\ codes by induction over $c=\ecol-\erow$.

 \begin{figure}
 \centering
     \resizebox{.65\totalheight}{!}{\input{EquivGraph.tex}}
     \caption{A flowchart of the proof of Theorem~\ref{theorem:equiv}. Given an array $\bfC_{\twot+1} \in \dball_{\erow,\erow+1}(\bfX_1) \cap \dball_{\erow,\erow+1}(\bfX_2)$, we show that there exists an array $\bfG_{\twot+1} \in \iball_{\erow,\erow+1}(\bfX_1) \cap \iball_{\erow,\erow+1}(\bfX_2)$. The series of orange marked arrays are given by the first assumption. Out of these we can proof the existence of the series of green marked arrays by Lemma~\ref{lemma:equivsimpl}~and~\ref{lemma:equiv}. The brown marked arrays are then given by applying Theorem~\ref{thm:tinsdel}. Lastly, the existence of the purple arrays can be shown again by Lemma~\ref{lemma:equivsimpl}~and~\ref{lemma:equiv}.}
     \label{fig:equivGraph}
 \end{figure}

\noindent \paragraph{Base case $c=1$} For the reader's convenience, a flowchart of the proof is presented in Figure~\ref{fig:equivGraph}.

Assume that there exists an array $\bfE \in \oi^{(n-\erow)\times(n-\erow-1)}$ such that $\bfE \in \dball_{\erow,\erow+1}(\bfX_1)\cap\dball_{\erow,\erow+1}(\bfX_2)$. Let $\twot = 2 \erow$, we define the series of arrays $\{\bfC_s\}_{s=1}^{\twot}$ to be the intermediate arrays obtained by deleting a row or a column from $\bfX_2$ to reach $\bfE$. For notational convenience we let $\bfC_0\triangleq \bfX_2$ and $\bfC_{\twot+1} \triangleq \bfE$. We follow an alternating order of deletion between columns and rows, i.e.,
\begin{equation*}
    \bfC_s \in \begin{cases}
    \dball_{0,1}(\bfC_{s-1}) & \hfill  \text{if $s$ is odd,}\\
    \dball_{1,0}(\bfC_{s-1}) & \hfill \text{ otherwise.}
    \end{cases}
\end{equation*}
Furthermore, we have $\bfC_{\twot+1}\in \dball_{0,1}(\bfC_{\twot})$. We denote by $\bfB_\twot \in \oi^{(n-\erow)\times(n-\erow)}$ the array resulting from deleting $\erow$ columns and $\erow$ rows from $\bfX_1$ such that $\bfE \in \dball_{0,1}(\bfB_\twot)$.

We now want to show that there exists a series of arrays $\{\bfB_s\}_{s=0}^{\twot-1}$ such that \begin{equation*}
    \bfB_s \in \begin{cases}
    \iball_{0,1}(\bfB_{s+1}) \cap \iball_{0,1}(\bfC_{s+1}) & \hfill \text{if $s$ is odd,}\\
    \iball_{1,0}(\bfB_{s+1}) \cap \iball_{0,1}(\bfC_{s+1}) & \hfill \text{otherwise.}\\
    \end{cases}
\end{equation*}
By definition, $\twot$ is even and $\bfC_{\twot+1} \in \dball_{0,1}(\bfB_\twot) \cap \dball_{0,1}(\bfC_\twot)$. By Lemma~\ref{lemma:equivsimpl} there exists an array $\bfB_{\twot-1} \in \oi^{(n-\erow) \times (n-\erow+1)}$ such that $\bfB_{\twot-1} \in \iball_{0,1}(\bfB_\twot) \cap \iball_{0,1}(\bfC_\twot)$. Moreover, we know that there exists $\bfC_{\twot-1} \in \oi ^{(n-\erow+1) \times (n-\erow)}$ such that $\bfC_{\twot}\in \dball_{1,0}(\bfC_{\twot-1})$. Consequently, we have that $\bfC_{\twot} \in \dball_{0,1}(\bfB_{\twot-1}) \cap \dball_{1,0}(\bfC_{\twot-1})$. By Lemma~\ref{lemma:equiv} there exists a $\bfB_{\twot - 2}\in \oi^{(n-\erow+1) \times (n-\erow+1)}$ such that $\bfB_{\twot-2} \in \iball_{1,0}(\bfB_{\twot-1}) \cap \iball_{0,1}(\bfC_{\twot-1})$.
Following the same arguments as above, one can show that for all even values of $s \in \{1,\dots,k\}$, the arrays $\bfC_s$, $\bfB_s$ and $\bfC_{s+1}$ satisfy $\bfC_{s+1} \in \dball_{0,1}(\bfC_s) \cap \dball_{0,1}(\bfB_s)$. Thus, by Lemma~\ref{lemma:equivsimpl} there exists an array $\bfB_{s-1} \in \iball_{0,1}(\bfC_s) \cap \iball_{0,1}(\bfB_s)$. Similarly, for all odd values of $s\in \{1,\dots,k\}$, the arrays $\bfC_s$, $\bfB_s$ and $\bfC_{s+1}$ satisfy $\bfC_{s+1} \in \dball_{1,0}(\bfC_s) \cap \dball_{0,1}(\bfB_s)$. Thus, by Lemma~\ref{lemma:equiv} there exists an array $\bfB_{s-1} \in \iball_{1,0}(\bfC_s) \cap \iball_{0,1}(\bfB_s)$. Subsequently, we will end up with an array $\bfB_0 \in \oi ^{n \times n}$ such that $\bfB_{\twot} \in \dball_{\erow,\erow}(\bfX_1) \cap \dball_{\erow,\erow}(\bfB_0)$. Therefore, by Theorem~\ref{thm:tinsdel} there exists an array $\bfF_{\twot} \in \oi ^{(n+\erow)\times(n+\erow)}$ such that $\bfF_\twot \in \iball_{\erow,\erow}(\bfX_1) \cap \iball_{\erow,\erow}(\bfB_0)$. 

Let $\bfF_0 \triangleq \bfB_0$ and $\bfF_{-1}\triangleq \bfC_1$, we denote again the series of arrays $\{\bfF_s\}_{s=1}^\twot$ that are the intermediate arrays obtained by a row or a column insertion starting from $\bfF_0$ until reaching $\bfF_\twot$. We again consider alternating insertions of rows and columns, i.e., 
\begin{equation*}
    \bfF_s \in \begin{cases}
        \iball_{1,0}(\bfF_{s-1}) & \hfill \text {if $s$ is odd,}\\
         \iball_{0,1}(\bfF_{s-1}) & \hfill \text {otherwise.}
    \end{cases}
\end{equation*}
Define the array $\bfG_0 \triangleq \bfX_2$, we will now show the existence of the series of arrays $\{\bfG_s\}_{s=1}^{\twot+1}$ such that
\begin{equation*}
    \bfG_s \in \begin{cases}
      \iball_{0,1}(\bfG_{s-1})\cap \iball_{0,1}(\bfF_{s-1}) & \hfill \text {if $s$ is odd,}\\
    \iball_{1,0}(\bfG_{s-1})\cap \iball_{0,1}(\bfF_{s-1}) & \hfill \text {otherwise.}
    \end{cases}
\end{equation*}
Following the same arguments used to construct the series $\{\bfB_s\}_{s=0}^{\twot-1}$, one can show that for all even values of $s \in \{0,\dots,k\}$, the arrays $\bfF_s$, $\bfG_s$ and $\bfF_{s-1}$ satisfy $\bfF_{s-1} \in \dball_{0,1}(\bfF_s) \cap \dball_{0,1}(\bfG_s)$. Thus, by Lemma~\ref{lemma:equivsimpl} there exists an array $\bfG_{s+1} \in \iball_{0,1}(\bfF_s) \cap \iball_{0,1}(\bfG_s)$. Similarly, for all odd values of $s\in \{0,\dots,\twot\}$, the arrays $\bfF_s$, $\bfG_s$ and $\bfF_{s-1}$ satisfy $\bfF_{s-1} \in \dball_{1,0}(\bfF_s) \cap \dball_{0,1}(\bfG_s)$. Thus, by Lemma~\ref{lemma:equiv} there exists an array $\bfG_{s+1} \in \iball_{1,0}(\bfG_s) \cap \iball_{0,1}(\bfF_s)$. As a consequence, we have shown that if there exists an array $\bfC_{\twot+1} \in \dball_{\erow, \erow+1}(\bfX_1) \cap \dball_{\erow, \erow+1} (\bfX_2)$, then there exists an array $\bfG_{\twot+1} \in \iball_{\erow, \erow+1}(\bfX_1)\cap \iball_{\erow,\erow+1}(\bfX_2)$.

\noindent \paragraph{Induction hypothesis} For any integer $c\geq 1$ it holds that for any $\bfX_1,\bfX_2\in \sigmatn$, $\dball_{\erow,\erow+c}(\bfX_1)\cap \dball_{\erow,\erow+c}(\bfX_2)\neq \emptyset$ if and only if $\iball_{\erow,\erow+1}(\bfX_1)\cap \iball_{\erow,\erow+c}(\bfX_2)\neq \emptyset$.

\noindent \paragraph{Induction step}

Assume that the induction hypothesis holds for all values $0\leq \ecol-\erow \leq c$. We prove that the hypothesis holds for $\ecol-\erow=c+1$.

Assume that there exists an array $\bfE \in \oi^{(n-\erow)\times(n-\erow-c-1)}$ such that $\bfE \in \dball_{\erow,\erow+c+1}(\bfX_1)\cap\dball_{\erow,\erow+c+1}(\bfX_2)$. Let $\twot= 2\erow$, define $\bfC_0 \triangleq \bfX_2$ and $\bfC_{\twot+c+1}\triangleq \bfE$. We denote by $\{\bfC_s\}_{s=0}^{\twot+c+1}$ the series of arrays resulting from a deletion of a row or a column starting from $\bfX_2$ until obtaining $\bfE$ as follows
\begin{equation*}
    \bfC_s \in \begin{cases}
        \dball_{0,1}(\bfC_{s-1}) & \text{if $s$ is odd or $s> \twot$,}\\
        \dball_{1,0}(\bfC_{s-1}) & \text{otherwise.}
    \end{cases}
\end{equation*}
We denote by $\bfB_{\twot+c} \in \oi^{(n-\erow)\times(n-\erow-c)}$ the array resulting from deleting $\erow$ columns and $\erow+c$ rows from $\bfX_1$ such that $\bfE \in \dball_{0,1}(\bfB_{\twot+c})$. We now want to show that there exists a series of arrays $\{\bfB_s\}_{s=0}^{\twot+c-1}$ such that \begin{equation*}
    \bfB_s \in \begin{cases}
    \iball_{0,1}(\bfB_{s+1}) \cap \iball_{0,1}(\bfC_{s+1}) & \text{if $s$ is odd or $s\geq \twot$,}\\
    \iball_{1,0}(\bfB_{s+1}) \cap \iball_{0,1}(\bfC_{s+1}) & \text{otherwise.}\\
    \end{cases}
\end{equation*}
Following the same arguments as in the base case, one can show that for all even values of $s \in \{0,\dots,\twot\}$ and for all values of $\twot \leq s\leq \twot+c$, the arrays $\bfC_s$, $\bfB_s$ and $\bfC_{s+1}$ satisfy $\bfC_{s+1} \in \dball_{0,1}(\bfC_s) \cap \dball_{0,1}(\bfB_s)$. Thus, by Lemma~\ref{lemma:equivsimpl} there exists an array $\bfB_{s-1} \in \iball_{0,1}(\bfC_s) \cap \iball_{0,1}(\bfB_s)$. Similarly, for all odd values of $s\in \{0,\dots,k\}$, the arrays $\bfC_s$, $\bfB_s$ and $\bfC_{s+1}$ satisfy $\bfC_{s+1} \in \dball_{1,0}(\bfC_s) \cap \dball_{0,1}(\bfB_s)$. Thus, by Lemma~\ref{lemma:equiv} there exists an array $\bfB_{s-1} \in \iball_{1,0}(\bfC_s) \cap \iball_{0,1}(\bfB_s)$. Subsequently, we will end up with an array $\bfB_0 \in \oi ^{n \times n}$ such that $\bfB_{\twot+c} \in \dball_{\erow,\erow+c}(\bfX_1) \cap \dball _{\erow,\erow+c}(\bfB_0)$. Therefore, by using the induction hypothesis, we can show the existence of the array $\bfF_{\twot+c} \in \oi^{(n+\erow)\times(n+\erow+c)}$ such that $\bfF_{\twot+c} \in \iball_{\erow,\erow+c}(\bfX_1) \cap \iball_{\erow,\erow+c}(\bfB_0)$.

Let $\bfF_0 \triangleq \bfB_0$ and $\bfF_{-1}\triangleq \bfC_1$, we denote again the series of arrays $\{\bfF_s\}_{s=1}^{\twot+c}$ as the intermediate arrays obtained by a row or a column insertion starting from $\bfF_0$ until reaching $\bfF_\twot$ as follows
\begin{equation*}
    \bfF_s \in \begin{cases}
        \iball_{0,1}(\bfF_{s-1}) & \text {if $s$ is even or $s>k$,}\\
         \iball_{1,0}(\bfF_{s-1}) & \text {otherwise.}
    \end{cases}
\end{equation*}
Define the array $\bfG_0 \triangleq \bfX_2$, we will now show the existence of the series of arrays $\{\bfG_s\}_{s=1}^{\twot+1}$ such that
\begin{equation*}
    \bfG_s \in \begin{cases}
      \iball_{0,1}(\bfG_{s-1})\cap \iball_{0,1}(\bfF_{s-1}) & \text {if $s$ is odd or $s> k$,}\\
    \iball_{1,0}(\bfG_{s-1})\cap \iball_{0,1}(\bfF_{s-1}) & \text {otherwise.}
    \end{cases}
\end{equation*}
Following the same arguments as in the base case, one can show that for all even values of $s \in \{0,\dots,\twot\}$ and $ \twot \leq s\leq \twot+c$, the arrays $\bfF_s$, $\bfG_s$ and $\bfF_{s-1}$ satisfy $\bfF_{s-1} \in \dball_{0,1}(\bfF_s) \cap \dball_{0,1}(\bfG_s)$. Thus, by Lemma~\ref{lemma:equivsimpl} there exists an array $\bfG_{s+1} \in \iball_{0,1}(\bfF_s) \cap \iball_{0,1}(\bfG_s)$. Similarly, for all odd values of $s\in \{0,\dots,\twot\}$, the arrays $\bfF_s$, $\bfG_s$ and $\bfF_{s-1}$ satisfy $\bfF_{s-1} \in \dball_{1,0}(\bfF_s) \cap \dball_{0,1}(\bfG_s)$. Thus, by Lemma~\ref{lemma:equiv} there exists an array $\bfG_{s+1} \in \iball_{1,0}(\bfG_s) \cap \iball_{0,1}(\bfF_s)$. As a consequence, we have shown that if there exists an array $\bfC_{\twot+c+1} \in \dball_{\erow, \erow+c+1}(\bfX_1) \cap \dball_{\erow, \erow+c+1} (\bfX_2)$, then there exists an array $\bfG_{\twot+c+1} \in \iball_{\erow, \erow+c+1}(\bfX_1)\cap \iball_{\erow,\erow+c+1}(\bfX_2)$. This concludes the induction.
\end{IEEEproof}

%% file: EquivGraph.tex
\newcommand{\LT}{L\ref{lemma:equivsimpl}}
\newcommand{\LI}{L\ref{lemma:equiv}}
\newcommand{\THRM}{Thm.~\ref{thm:tinsdel}}

\begin{tikzpicture}[
    arr/.style = {rectangle, rounded corners, draw=black,
                           minimum width=10ex, minimum height=2ex,
                           text centered, font=\normalsize},
    givenbefore/.style = {arr, fill=color1!30},
    createdfirst/.style = {arr, fill=color2!30},
    giventhen/.style = {arr, fill=color6!30},
    createdsecond/.style = {arr, fill=color4!30},
    thm/.style = {circle, draw=black, fill=color3!30,
                           text centered, font=\normalsize},
    arrowdescp/.style = {midway, fill=white, font=\footnotesize}]

    \def\xdist{8.2ex}
    \def\ydist{8.2ex}
    
    \definecolor{color0}{rgb}{0.12156862745098,0.466666666666667,0.705882352941177}
    \definecolor{color1}{rgb}{1,0.498039215686275,0.0549019607843137}
    \definecolor{color2}{rgb}{0.172549019607843,0.627450980392157,0.172549019607843}
    \definecolor{color3}{rgb}{0.83921568627451,0.152941176470588,0.156862745098039}
    \definecolor{color4}{rgb}{0.580392156862745,0.403921568627451,0.741176470588235}
    \definecolor{color6}{rgb}{0.549019607843137,0.337254901960784,0.294117647058824}
    \definecolor{color5}{rgb}{0.890196078431372,0.466666666666667,0.76078431372549}
    
    \node [arr,fill=color0!30] (x1) at (0,0) {$\bfX_1$};
    
    \node [giventhen] (fk) at (0*\xdist,4*\ydist) {$\bfF_\twot$};
    \node [createdsecond] (gk) at (2*\xdist,4*\ydist) {$\bfG_{\twot}$};
    \node [givenbefore] (ck) at (2*\xdist,-4*\ydist) {$\bfC_{\twot}$};
    \node [givenbefore] (bk) at (0*\xdist,-4*\ydist) {$\bfB_\twot$};
    
    \node [arr,fill=color0!30] (gkp1) at (1*\xdist,5*\ydist) {$\bfG_{\twot+1}$};
    \node [arr,fill=color0!30,] (ckp1) at (1*\xdist,-5*\ydist) {$\bfC_{\twot+1}$};
    
    \node [giventhen] (fkm1) at (1*\xdist,3*\ydist) {$\bfF_{\twot-1}$};
    \node [createdsecond] (gkm1) at (3*\xdist,3*\ydist) {$\bfG_{\twot-1}$};
    \node [givenbefore] (ckm1) at (3*\xdist,-3*\ydist) {$\bfC_{\twot-1}$};
    \node [createdfirst] (bkm1) at (1*\xdist,-3*\ydist) {$\bfB_{\twot-1}$};
    
    \node [giventhen] (fkm2) at (2*\xdist,2*\ydist) {$\bfF_{\twot-2}$};
    \node [createdsecond] (g2) at (4*\xdist,2*\ydist) {$\bfG_{2}$};
    \node [givenbefore] (c2) at (4*\xdist,-2*\ydist) {$\bfC_{2}$};
    \node [createdfirst] (bkm2) at (2*\xdist,-2*\ydist) {$\bfB_{\twot-2}$};
    
    \node [giventhen] (f1) at (3*\xdist,1*\ydist) {$\bfF_{1}$};
    \node [createdsecond] (g1) at (5*\xdist,1*\ydist) {$\bfG_{1}$};
    \node [givenbefore] (c1) at (5*\xdist,-1*\ydist) {$\bfC_{1}$};
    \node [createdfirst] (b1) at (3*\xdist,-1*\ydist) {$\bfB_{1}$};
    
    \node [createdfirst] (f0) at (4*\xdist,0) {$\bfB_0 = \bfF_0$};
    \node [arr,fill=color0!30] (g0) at (6*\xdist,0) {$\bfX_{2}$};
    
    \draw [-] (x1) -- (fk) node [arrowdescp] {\erow rows, \erow columns};
    \draw [-] (x1) -- (bk) node [arrowdescp] {\erow rows, \erow columns};
    
    \draw [-] (fk) -- (gkp1) node [arrowdescp] {column};
    \draw [-] (gk) -- (gkp1) node [arrowdescp] {column};
    \draw [-] (gk) -- (fkm1) node [arrowdescp] {column};
    \draw [-] (fk) -- (fkm1) node [arrowdescp] {column};
    \node [thm] (uk) at (1*\xdist,4*\ydist) {\LT};
    
    \draw [-] (fkm1) -- (fkm2) node [arrowdescp] {row};
    \draw [-] (gk) -- (gkm1) node [arrowdescp] {row};
    \draw [-] (fkm2) -- (gkm1) node [arrowdescp] {column};
    \node [thm] (ukm1) at (2*\xdist,3*\ydist) {\LI};
    
    \draw [dashed] (gkm1) -- (g2);
    \draw [dashed] (fkm2) -- (f1);
    
    \draw [-] (f1) -- (f0) node [arrowdescp] {row};
    \draw [-] (f1) -- (g2) node [arrowdescp] {column};
    \draw [-] (g1) -- (g2) node [arrowdescp] {row};
    \draw [-] (f0) -- (g1) node [arrowdescp] {column};
    \node [thm] (u1) at (4*\xdist,1*\ydist) {\LI};
    
    \draw [-] (g0) -- (g1) node [arrowdescp] {column};
    \draw [-] (g0) -- (c1) node [arrowdescp] {column};
    \draw [-] (f0) -- (c1) node [arrowdescp] {column};
    \node [thm] (ul0) at (5*\xdist,0*\ydist) {\LT};
    
    \draw [-] (f0) -- (b1) node [arrowdescp] {row};
    \draw [-] (g0) -- (c1) node [arrowdescp] {column};
    \draw [-] (c1) -- (c2) node [arrowdescp] {row};
    \draw [-] (b1) -- (c2) node [arrowdescp] {column};
    \node [thm] (l1) at (4*\xdist,-1*\ydist) {\LI};
    
    \draw [dashed] (bkm2) -- (b1);
    \draw [dashed] (ckm1) -- (c2);
    
    \draw [-] (bkm1) -- (bkm2) node [arrowdescp] {row};
    \draw [-] (ck) -- (ckm1) node [arrowdescp] {row};
    \draw [-] (bkm2) -- (ckm1) node [arrowdescp] {column};
    \draw [-] (bkm1) -- (ck) node [arrowdescp] {column};
    \node [thm] (lkm1) at (2*\xdist,-3*\ydist) {\LI};
    
    \draw [-] (bk) -- (bkm1) node [arrowdescp] {column};
    \draw [-] (ck) -- (ckp1) node [arrowdescp] {column};
    \draw [-] (bk) -- (ckp1) node [arrowdescp] {column};
    \node [thm] (lk) at (1*\xdist,-4*\ydist) {\LT};
    
    \node [thm] (lk) at (1.75*\xdist,0*\ydist) {\THRM};

\end{tikzpicture}

%% file: card-bound_loglog.tex
This section presents an asymptotic upper bound on the cardinality of any \tcc. It implies an asymptotic lower bound on the redundancy of any binary \tcc, denoted by \redopt.

\begin{lem}\label{lem:card-bound}
Any upper bound on the cardinality of a $q$-ary $\etot$-deletion-correcting code $\cC_{q,n,\etot}$ with $q=2^n$ is also an upper bound on the cardinality of a binary \tcc. 
\end{lem}
\begin{IEEEproof}
Note that a $2^n$-ary $\etot$-deletion-correcting code $\cC_{2^n,n,\etot}$ can be seen also as a binary $\etot$ column deletion-correcting code by interpreting the symbols as binary columns. Since a \tcc\ $\cC$ can correct any combination of $\erow$ row and $\ecol$ column deletions such that $\erow + \ecol = \etot$, in particular it can also correct any $\etot$ column deletions. Therefore, any upper bound on the size of $\cC_{2^n,n,\etot}$ is also a valid upper bound on the size of $\cC$.
\end{IEEEproof}

\begin{cor}
For any binary \tcc\ $\cC$ it holds that
\begin{equation*}
    \lvert \cC \rvert \lesssim \frac{\etot ! 2^{n^2}}{(2^n-1)^\etot n^\etot}.
\end{equation*}
Consequently, we have $\redopt \gtrsim \etot n + \etot \log (n) - \log(\etot !) $.
\end{cor}
\begin{IEEEproof}
From~\cite{Kulkarni_Nonasymptotic-Upper-Bounds-for-Deletion-Correcting-Codes_13}, we have for a $q$-ary $\etot$-deletion correcting code that $\lvert \cC_{q,n,\etot} \rvert \lesssim \frac{\etot ! q^n}{(q-1)^\etot n^\etot}$ when $q$ is fixed. By following the same arguments as in the proof of \cite[Theorem 4.3]{Kulkarni_Nonasymptotic-Upper-Bounds-for-Deletion-Correcting-Codes_13}[Theorem 4.3] we can show that this bound holds also true when $q=2^n$.
Therefore, for any binary \tcc\ $\cC$ it holds by Lemma~\ref{lem:card-bound} that
\begin{align*}
    \lvert \cC \rvert \leq \lvert \cC_{2^n,n,\etot} \rvert  \lesssim \frac{\etot ! 2^{n^2}}{ (2^n-1)^\etot n^\etot}.
\end{align*}
Therefore, we have
\begin{align*}
    \redopt \gtrsim n^2 - \log(|\cC|)  
       \approx \etot n + \etot \log (n) - \log(\etot !).
\end{align*}
\end{IEEEproof}

%% file: construction_loglog.tex
In this section we present an existential construction of \tccs. We start with an intuitive road map to our code construction and then formally define each ingredient.
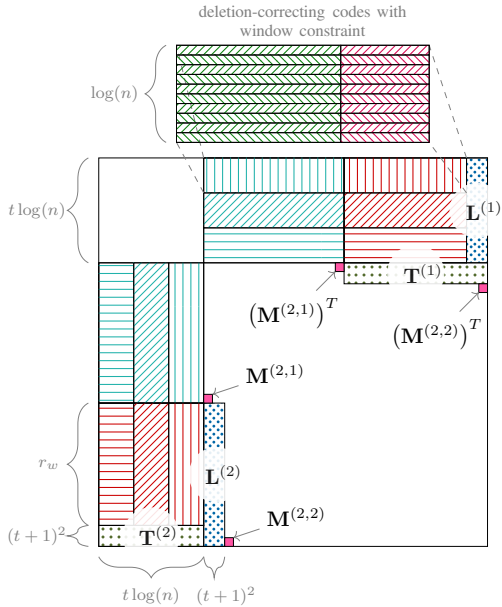
\begin{figure}[b!]
    \centering
    \resizebox{7cm}{!}{
    \input{tikz_cons_loglog}
    }
    \caption{Illustration of an array contained in the locator set $\locaset$ for $\etot = 3$. In the first $\etot \log(n)$ rows there are $\etot$ blocks each consisting of a systematic part (\textcolor{lightseagreen}{cyan}) and a redundancy part (\textcolor{bostonuniversityred}{red}). Each row is encoded using a systematic $\etot$-deletion-correcting code (zoomed in part). In addition, in the systematic part of each block a window constrained is imposed. Those blocks are used to locate column deletions. This structure is protected with the arrays $\leftmargo^{(1)}$ (\textcolor{mediumpersianblue}{blue}) against row deletions and $\topmargo^{(1)}$ (\textcolor{darkolivegreen}{brown}) against column deletions. Lastly, to locate the borders of $\topmargo^{(1)}$ we use the marker arrays $\econstarr{2,1}$ and $\econstarr{2,2}$ (\textcolor{brilliantrose}{pink}). A symmetric structure locates row deletions.
    }
    \label{fig:cons-loglog}
\end{figure}
\subsection{Road Map}
Our construction uses structured arrays so that the indices of the inserted/deleted rows and columns can be exactly recovered. Then, the set of structured arrays is intersected with arrays of a Gabidulin code (that can correct row/column erasures) to recover the arrays of the code. The structure is depicted in Figure~\ref{fig:cons-loglog}.

We structure the $n\times n$ codewords $\bfC$ as follows. We protect the columns with indices between $\etot \log n +1$ and $n-(t+1)^2$ using $\etot \log n$ codes where each one is a binary systematic $\etot$-\insdel-correcting code. We divide those codes into $t$ blocks each of size $\log n$. We impose what we call a \emph{window constraint} on the columns of the systematic part of every block. This constraint ensures that every $\etot+1$ consecutive columns are different. Therefore, the indices of the deleted columns within the systematic part can be located by using all $\log n$ \insdel-correcting codes of any block (Claim~\ref{cl:deletion-locate-set}). In case of insertions, the indices of the inserted columns within the systematic part can be located up to an interval of length at most $2\etot$ containing at most $\etot$ columns of the original array. This ambiguity arises when the inserted rows/columns are equal to collections of rows/columns of the original array within an interval. We call this phenomenon \emph{block confusions}  (cf. Section~\ref{sec:locator-arrays} and Claim~\ref{cl:insertion-locate-set}).

In the redundancy part, runs may exist. Thus, the recovery of the index of the deleted columns is only guaranteed within the corresponding run. To recover the exact location of the deleted columns here, we protect the redundancy part of the codes by appending (from below) what we call a \emph{locator array} that can detect the exact positions of column deletions within this part (Claim~\ref{cl:leftmargo-deletions}). We call this array $\topmargo^{(1)}$. In case of insertions, the locator array is used to recover the index of the inserted columns up to block confusions of length at most $2\etot$ (Claim~\ref{cl:leftmargo-insertions}).

Note that for the window constraint to work, we need to have all $\log n$ \insdel-correcting codes of the considered block. Therefore, we use the subarray $\bfC_{[1:\etot\log n],[n-(t+1)^2:n]}$ as a locator array $\leftmargo^{(1)}$ that can detect the exact position of a deleted row within the first $\etot\log n$ rows (Claim~\ref{cl:leftmargo-deletions}). As a result, if all $\etot$ deletions are row deletions within the first $\etot \log n$ rows, then the locator array is enough to recover all the indices (Lemma~\ref{lem:index-recover-manab-del}). Otherwise, we have at least one block of the $\etot$ blocks that is not affected by a row deletion. This block is used to recover the deleted columns with indices in the range $\etot \log n +1$ to $n-(t+1)^2$ (Lemma~\ref{lem:index-recover-delcode-del}). The same arguments hold for the insertion case (Claim~\ref{cl:leftmargo-insertions}, Lemma~\ref{lem:index-recover-manab-insertion}~and Lemma~\ref{lem:index-recover-delcode-insertion}). The only difference is the possibility of the inserted row/column causing ambiguities for the exact indices of the insertions.

One more step is needed. We must be able to locate the position of the locator arrays within the resulting $(n-\erow) \times (n-\ecol)$ array $\widetilde{\bfC}$. Therefore, we put four \emph{marker arrays} after the locators that are detectable even after $\etot$ insertions or deletions. We call those arrays $\econstarr{1,1}$ and $\econstarr{1,2}$.

The same structure (transposed) is used to index the rows. In addition, the columns with indices between $1$ and $\etot \log n$ are protected by the locator array used for protecting the \insdel-correcting codes indexing the rows. Note the claims and lemmas mentioned before also include the statements to recover the row indices.

The whole code is intersected with a Gabidulin code~\cite{Gabidulin_TheoryOfCodes_1985} that can correct row/column erasures. Once the positions of the deleted rows and columns are known to the decoder, those positions are marked as erasures and corrected using the Gabidulin code. In case of insertions, all inserted rows and columns with exactly recovered indices are removed. In the case where inserted rows/columns create block confusions we can simply delete all involved rows/columns and insert erasures. The window constraint guarantees that we do not delete more than $\etot$ rows/columns of the original array by this procedure, hence assuring to not exceed the erasure correction capability of the Gabidulin code. 

In the next subsections, we formally define the five main ingredients of our code: \begin{enumerate*}[label=\textit{(\roman*)}]\item the locator arrays; \item the binary systematic $\etot$-deletion-correcting codes with window constraints; \item the marker arrays; \item the \emph{locator set} which is the combination of all the previously mentioned parts; and \item a Gabidulin code~\cite{Gabidulin_TheoryOfCodes_1985} that is used to correct row/column erasures.\end{enumerate*}

\subsection{Locator Arrays}\label{sec:locator-arrays}

For a positive integer $\a$, we denote by \idmat{\a} the identity array of dimension $\a \times \a$ and by \allone{\a} and \allzero{\a} the all-one row vector and all-zero row vector of length $\a$, respectively. 
We use $\otimes$ to indicate the Kronecker product.
We thus have the following definition from \cite{manabu2020delarrays}.

\begin{definition}[Locator arrays] \label{def:manabu-leftmarg}
We set $\leftm{\prime} \in \Sigma ^{(t+1) \times (t+1)^2}$ as $\leftm{\prime} \triangleq  \idmat{\etot+1} \otimes \allone{\etot+1} $. More precisely, $\leftm{\prime}$ has the following structure
\begin{align*}
\leftm{\prime} = 
\begin{pmatrix}
\allone{\etot+1} & \allzero{\etot+1} & \dots & \allzero{\etot+1}\\
\allzero{\etot+1} & \allone{\etot+1} & \dots & \allzero{\etot+1}\\
\vdots & \vdots& \ddots & \vdots\\
\allzero{\etot+1} & \allzero{\etot+1} & \dots & \allone{\etot+1} \\
\end{pmatrix}.
\end{align*}
Let $\s$ be a multiple of $(\etot+1)$ such that $\s \geq \left\lceil \frac{\etot}{2} \rceil\right (\etot+1)$. We define the locator array $\leftmarg{\s} \in \Sigma^{\s \times (\etot+1)^2}$ as 
\begin{equation*}
    \leftmarg{\s} \triangleq \allone{\frac{\s}{\etot+1}} ^T \otimes \leftm{\prime} .
\end{equation*}
Moreover, we define the locator array $\topmarg{\s} \in \Sigma^{ (\etot+1)^2 \times \s}$ to be the transpose of $\leftmarg{\s}$, i.e., 
\begin{equation*}
    \topmarg{\s} \triangleq \leftmarg{\s}^T = \allone{\frac{\s}{\etot+1}} \otimes \leftm{\prime \, T}  .
\end{equation*}
\end{definition}
Throughout the paper we drop $\s$ in the notation $\leftmarg{\s}$ and $\topmarg{\s}$ when the value of $s$ is clear from the context.

\begin{claim}[Deletion detection in $\leftmarg{s}$ and $\topmarg{s}$]\label{cl:leftmargo-deletions}
Let $\leftmarg{\s}$ be an array affected by $\erow$ row and $\ecol$ column deletions such that $\erow+\ecol = \etot$. Divide $\leftmarg{\s}$ into $(\etot+1)$ subarrays each consisting of $(\etot+1)$ consecutive columns of $\leftmarg{s}$. By examining $\widetilde{\leftmargo}_{\s}$, we can locate the exact positions of the deleted rows. We can also determine the number of column deletions that happened in each subarray of $\leftmarg{\s}$.

Let $\topmarg{\s}$ be an array affected by $\erow$ row and $\ecol$ column deletions such that $\erow+\ecol = \etot$. The same statement above for $\leftmarg{\s}$ holds for $\topmarg{\s}$ by switching rows for columns. 
\end{claim}
\begin{IEEEproof}
We prove the first part of the claim, while the second part follows similarly since $\topmarg{\s} = \leftmarg{\s}^T$ and row deletions in $\topmarg{\s}$ can be seen as column deletions in $\leftmarg{\s}$ and vice versa.

By construction of $\leftmarg{\s}$, for any $i \in [\s-\etot]$ and $j \in [\etot]$  it holds that $\leftmargo_{i,[(\etot+1)^2]} \neq \leftmargo_{i+j,[(\etot+1)^2]}$. This property holds true even in the presence of at most $\etot$ column deletions in $\leftmarg{\s}$. Thus, due to the fixed structure of $\leftmarg{\s}$ one can uniquely determine the exact indices of the deleted rows.

Moreover, we divide $\leftmarg{\s}$ in subarrays consisting of ${(\etot +1)}$ columns. For any $a \in [\etot+1]$ and $b \in [\etot+1]$, we have $\leftmargo _{[\s],(a-1)(\etot+1)+1} = \leftmargo _{[\s],(a-1)(\etot+1)+b} $. In words, we have $(\etot+1)$ identical columns in a subarray. This property holds true even if there were at most $\etot$ row deletions in $\leftmarg{\s}$. Furthermore, for $\s \geq \left\lceil \frac{\etot}{2} \rceil\right (\etot+1)$ and $a,b,c \in [\etot+1]$, it always holds that $\leftmargo _{[\s],(a-1)(\etot+1)+1} \neq \leftmargo _{[\s],(c-1)(\etot+1)+b}$, unless $a=c$. Therefore, we can determine the deleted columns within any subarray by counting the number of missing columns.  
\end{IEEEproof}

We briefly elaborate on the dimension constraint of $\leftmarg{s}$, \ie $s \geq \left\lceil \frac{\etot}{2} \rceil\right (\etot+1)$. If $\mathbf{L}_s$ consists of less than $\left\lceil \frac{t}{2}\right\rceil$ copies of $\mathbf{L}^{\prime}$, then a deletion of two consecutive rows in each block will lead to an impossibility in locating a column deletion within two adjacent subarrays. Recall that $\allone{a}$ is the all-one vector of length $a$ and let $\etot=3$ and $s=4$, we assume the following deletion pattern:
\begin{center}
    \begin{tikzpicture}

    \node (a) {$\begin{pmatrix}
    \mathbf{1}_4 & \mathbf{0}_4 & \mathbf{0}_4 & \mathbf{0}_4 \\
    \mathbf{0}_4 & \mathbf{1}_4 & \mathbf{0}_4 & \mathbf{0}_4 \\
    \mathbf{0}_4 & \mathbf{0}_4 & \mathbf{1}_4 & \mathbf{0}_4 \\
    \mathbf{0}_4 & \mathbf{0}_4 & \mathbf{0}_4 & \mathbf{1}_4 
    \end{pmatrix}$};
    
    \node  (b) [right of=a, node distance = 5.5cm] {$\begin{pmatrix}
    \mathbf{0}_7 & \mathbf{1}_4 & \mathbf{0}_4 \\
    \mathbf{0}_7 & \mathbf{0}_4 & \mathbf{1}_4 
    \end{pmatrix}$};
    
    \draw[->] (a.east) -- (b.west) node[midway,above] {\footnotesize 1st \& 2nd row del.} node[midway,below] {\footnotesize 2nd column del.};
    
    \draw [decorate,decoration={brace,amplitude=5pt},xshift=0pt,yshift=0pt]
(a.south east) -- (a.south west) node [black,midway,yshift=-0.5cm] 
{\footnotesize $\leftm{\prime}$};

\draw [decorate,decoration={brace,amplitude=5pt},xshift=0pt,yshift=0pt]
(b.south east) -- (b.south west) node [black,midway,yshift=-0.5cm] 
{\footnotesize $\widetilde{\mathbf{L}}^{\prime}$};
    
    \end{tikzpicture}
    
\end{center}
Even though we can locate the row deletions, we cannot locate the column deletion within a subarray of $(t+1)$ columns. Thus, it is necessary to have another copy of $\mathbf{L}^{\prime}$ within $\mathbf{L}_s$ to guarantee a successful detection of column deletions. In the general case, this extends to needing at least $\left\lceil \frac{t}{2}\right\rceil$ copies of $\mathbf{L}^{\prime}$ in $\mathbf{L}_s$. Note that this example can be directly applied for $\topmarg{s}$ by switching in the argument rows and columns.

For the following statements and proofs in case of insertions we have to define the terminology of block confusions.

\emph{Block confusions:}
Consider an array $\bfZ \in \Sigma^{n \times m}$. Let $\widetilde{\bfZ}$ be the resulting array after $\etot$-criss-cross insertions in $\bfZ$.
Given a subset $\mathcal{B} \subseteq [n]$, an integer $a \in [n]$, and a non-negative integer $c \leq n$, we define a row block confusion in $\widetilde{\bfZ}$ if $\widetilde{\bfZ}_{a+c',[m]} = \widetilde{\bfZ}_{a+b+c',[m]}$ for all $b \in \mathcal{B}$ and all non-negative $c'$ such that $c' \leq c$.

In other words, we declare a block confusion after insertions, when a collection of $c'$ consecutive rows in $\widetilde{\bfZ}$ are the same as other collections of consecutive rows within an interval determined by $a$ and $\mathcal{B}$. This actually leads to the fact that an insertion locating algorithm cannot determine the exact location (up to the block confusion) of the inserted rows even when knowing $\bfZ$.
For convenience, we provide the following illustration of a row block confusion.
\begin{center}
    \begin{tikzpicture}

    \node (a) {$\begin{pmatrix}
    1 & 0 & 1 & 0 \\
    0 & 1 & 0 & 1 
    \end{pmatrix}$};
    
    \node  (b) [right of=a, node distance = 5.5cm] {$\begin{pmatrix}
    1 & 0 & 1 & 0 \\
    0 & 1 & 0 & 1 \\
    1 & 0 & 1 & 0 \\
    0 & 1 & 0 & 1 
    \end{pmatrix}$};
    
    \draw[->] (a.east) -- (b.west) node[midway,above] {\footnotesize insertion in} node[midway,below] {\footnotesize 1st \& 2nd row};
    
    \draw [decorate,decoration={brace,amplitude=5pt},xshift=0pt,yshift=0pt]
(a.south east) -- (a.south west) node [black,midway,yshift=-0.5cm] 
{\footnotesize $\bfZ$};

\draw [decorate,decoration={brace,amplitude=5pt},xshift=0pt,yshift=0pt]
(b.south east) -- (b.south west) node [black,midway,yshift=-0.5cm] 
{\footnotesize $\widetilde{\bfZ}$};
    \end{tikzpicture}
\end{center}
Given both $\bfZ$ and $\widetilde{\bfZ}$, an insertion locating algorithm cannot distinguish whether the insertion happened in the first and second row or third and fourth row. We call this a row block confusion with parameters $a=1$, $\mathcal{B} = \{2\}$, and $c = 1$.

In the case that $\bfZ$ satisfies predetermined properties we can narrow down the parameter range of the possible block confusions. Let $\bfZ \in \Sigma ^{n \times m}$ be an array satisfying $\bfZ_{i,[m]} \neq \bfZ_{i+j,[m]}$ for all $i \in [n-\etot]$ and $j \in [\etot]$. Then, it follows that $\mathcal{B} \subseteq [\etot]$, $a \in [n]$, and $0 \leq c \leq \etot$. Therefore, the window of a row block confusion, defined as $[a,\max(\mathcal{B})+c]\triangleq\{a, \dots,\max(\mathcal{B})+c\}$, consists of at most $2\etot$ rows. In addition, within this window there exists at least one and at most $\etot$ rows of the original array $\bfZ$. Moreover, the bounds on the maximum number of original columns in the confusion and the \emph{length} of the confusion, defined as the total number of rows in the block confusion, scale proportionally with the number of insertions responsible for the confusion. Note that the aforementioned property on the array $\bfZ$ is satisfied by $\leftmarg{s}$.

We define a column block confusion similarly. The array $\topmarg{s}$ satisfies the desired properties on the columns to limit the parameter range of column block confusions. In the sequel, we will drop the row and column term when referring to block confusions when it is clear from the context.

\begin{claim}[Insertion detection in $\leftmarg{s}$ and $\topmarg{s}$]\label{cl:leftmargo-insertions}
Let $\leftmarg{\s}$ be affected by $\erow$ row and $\ecol$ column insertions such that $\erow+\ecol = \etot$. Divide $\leftmarg{\s}$ into $(\etot+1)$ subarrays each consisting of $(\etot+1)$ consecutive columns of $\leftmarg{s}$. By examining $\widetilde{\leftmargo}_{\s}$, we can locate the positions of the inserted rows up to a row block confusion of length at most $2\etot$ containing at most $\etot$ columns of the original array $\leftmarg{s}$. We can also determine the number of column insertions (and possibly their positions up to a column block confusion within the subarray or within the adjacent subarray) that happened in each subarray of $\leftmarg{\s}$.

Let $\topmarg{\s}$ be affected by $\erow$ row and $\ecol$ column insertions such that $\erow+\ecol = \etot$. The same statement above for $\leftmarg{\s}$ holds for $\topmarg{\s}$ by switching rows for columns. 
\end{claim}
\begin{IEEEproof}
The proof follows the same steps as in Claim~\ref{cl:leftmargo-deletions}. In terms of row insertions, the difference is that if the inserted rows in $\leftmarg{s}$ create a block confusion, then we cannot distinguish between the original rows and the inserted rows in a window of length at most $\etot +1$ rows. This follows from the construction of $\leftmarg{s}$. In terms of column insertions, if the inserted column is different from the column run in which it is inserted and from both adjacent column runs, then it can be exactly located. Otherwise, we can only count the number of insertions that happened in each block up to a confusion with an adjacent block. This is due to possible column block confusions at the column runs located in adjacent subarrays.
\end{IEEEproof}

\subsection{\Insdel-Correcting Codes with Window Constraints}

\emph{\Insdel-correcting codes:} We use the construction of \cite{sima2020systematictdel} for our binary systematic $\etot$-{\insdel}-correcting code. We briefly recall the results of~\cite{sima2020systematictdel}. Given a sequence $\bfk \in \Sigma ^{\k}$, one can compute a redundancy vector $\bfr_\bfk \in \Sigma ^{r_k}$ with $ r_k \leq 4 \etot \log (\k) + o(\log (\k))$. The resulting sequence $(\bfk | \bfr_\bfk)$  can be uniquely recovered after $\etot$ {\insdel} errors. Note that $\bfr_\bfk$ is a function of the information $\bfk$ and $\ro_\k$ is a function of the information length $\k$ and the number of {\insdel} errors $\etot$. 

\emph{Window constraint:} We define the \wcons\ as the set $\winarr{\ell,w} \subseteq \Sigma^{\ell \times w}$, where for any $\bfW \in \winarr{\ell,w}$, $i \in [w-\etot]$ and $j \in [\etot]$, it holds that $\bfW_{[\ell],i} \neq \bfW_{[\ell],i+j}$. 

For an array $\bfW \in \winarr{\ell,w}$, let $\bfR _\bfW \in \Sigma ^ {\ell \times r_w}$ be the array formed such that for any $i \in [\ell]$ the $i\th$ row of $\bfR _\bfW$ is the redundancy vector corresponding to the $i\th$ row of $\bfW$; computed using the systematic construction in~\cite{sima2020systematictdel}. We refer to the array $\bfR _\bfW \in \Sigma ^ {\ell \times r_w}$ as the redundancy array. Let $m \triangleq w + r_w$, we define $\edelarr{1}(\ell,m)$ as the set of all arrays resulting from the concatenation of $\bfW$ and $\bfR_\bfW$, i.e., 
\begin{align*}
 \edelarr{1}(\ell,m)  
  \triangleq \left\{\bfD \in \Sigma ^{\ell \times m} : 
  \begin{aligned}
  &{\bfD = (\bfW \mid \bfR_\bfW)}, \\
  &{\  \mathrm{s. t. } \bfW \in \winarr{\ell,w} }
  \end{aligned}\right\}.
\end{align*}
In words, $\edelarr{1}(\ell,m)$ is the set of binary systematic $\etot$-\insdel-correcting codes in which the systematic part satisfies the imposed window constraint. This set will be used to index the columns of our arrays in the constructed code. We define $\edelarr{2}(\ell,m)\triangleq \left\{\bfD^T: \bfD \in \edelarr{1}(\ell,m) \right\}.$ This set is going to be used for indexing the rows.
\begin{claim} \label{cl:deletion-locate-set}
Let $\bfD = (\bfW \mid \bfR_\bfW) \in  \edelarr{1}(\ell,m)$ be an array affected by $\etot$ column deletions and no row deletions, we can locate the exact positions of the deleted columns in the subarray $\bfW$. 

The same holds for any array in $\edelarr{2}(\ell,m)$ by switching in the argument rows and columns.
\end{claim}
\begin{IEEEproof}
Assume $\widetilde{\bfD} = (\widetilde{\bfW} \mid \widetilde{\bfR}_\bfW )$ is the array obtained after the deletions. For each row in $\widetilde{\bfW}$ we can use the corresponding redundancy in $\widetilde{\bfR} _ \bfW$ to correct the deletions that happened in this row~\cite{sima2020systematictdel}. We start by looking at the position of the first recovered bit in each row. In each row, this position may be unique or may be in an interval of possible positions (run). The exact location of the column is then determined by the unique position in which all runs (of all rows) intersect. The intersection is guaranteed to be unique by the imposed window constraint; since for any  $i \in [w-\etot]$, and $j \in [\etot]$, it holds that $\bfW_{[\ell],i} \neq \bfW_{[\ell],i+j}$. This process is repeated for all recovered bits until all $\etot$ positions are determined.

A similar argument follows for the second statement of the claim.
\end{IEEEproof}

\begin{claim} \label{cl:insertion-locate-set}
Given an array $\bfD = (\bfW \mid \bfR_\bfW) \in  \edelarr{1}(\ell,m)$ affected by $\etot$ column insertions and no row insertions, we can locate the positions of the inserted columns in the subarray $\bfW$ up to column block confusion of length at most $2\etot+1$ containing at most $\etot$ columns of the original array $\bfD$. 

The same holds for any array in $\edelarr{2}(\ell,m)$ by switching in the argument rows and columns.
\end{claim}
\begin{IEEEproof}
The proof follows the same technique of the proof of Claim~\ref{cl:deletion-locate-set}. The only exception arises if the inserted columns create a column block confusion. However, the window constraint guarantees that in the worst case the block confusion occurs in a window of length at most $2\etot$. Moreover, within this block confusion there exists at most $\etot$ columns of the original array. 
\end{IEEEproof}

\subsection{Marker Arrays}

We define the following arrays of dimension $(\etot+1) \times (\etot+1)$ which will operate as \emph{markers} to locate the position of the locator arrays in the resulting $\widetilde{\bfC}$. Recall that we use four locator arrays in our construction, namely $\leftmargo^{(1)}$, $\leftmargo^{(2)}$, $\topmargo^{(1)}$, and $\topmargo^{(2)}$, cf. Figure~\ref{fig:cons-loglog}. We only need marker arrays for $\topmargo^{(1)}$ and $\leftmargo^{(2)}$. The position of $\leftmargo^{(1)}$ and $\topmargo^{(2)}$ can be then determined. The first marker array $\econstarr{2,1}$, put on top of $\leftmargo^{(2)}$, consists of the first $\etot+1$ columns of $\leftmargo'$. The second marker array $\econstarr{2,2}$, put on the right of $\leftmargo^{(2)}$, consists of the complement of the last $\etot+1$ columns of $\leftmargo'$. The marker arrays $\econstarr{1,1}$ and $\econstarr{1,2}$ are the transpose of $\econstarr{2,1}$ and $\econstarr{2,2}$, respectively.

\subsection{Locator Set}
We formally define the sets of arrays in $\sigmatn$ that form our code. Let $\bfX \in \sigmatn$, we start with the set of arrays that are used to index the columns. This set is denoted by $\horzedel{\ell,n}$. The arrays in this set have the first $\etot\ell$ columns divided into $\etot$ blocks. The columns whose indices are between $\etot\ell+1$ and $n-(\etot+1)^2$ of each row consist of a systematic $\etot$-deletion-correcting code in which the systematic part satisfies the window constraint. We can write
\begin{align*}
  \horzedel{\ell,n}
  &\triangleq\!\left\{\bfX\!: \!
  \begin{aligned}
  &{\bfX_{[(a-1)\ell+1:a\ell],[\etot \ell+1:n-(\etot+1)^2]}} \\
  &{ \quad \in \edelarr{1}(\ell,n-\etot \ell - (\etot+1)^2) \ \forall \, a \in [\etot]}  
  \end{aligned}\right\}.
\end{align*}
The set of arrays $\vertedel{\ell,n}$ that are used to index the rows is defined similarly to $\horzedel{\ell,n}$ by replacing columns with rows
\begin{align*}
     \vertedel{\ell,n}
     &\triangleq\! \left\{\bfX\!:\!
     \begin{aligned}
    &{\bfX_{[\etot \ell+1:n-(\etot+1)^2],[(b-1)\ell+1:b\ell]} }\\
    &{ \quad \in \edelarr{2}(\ell,n-\etot \ell - (\etot+1)^2) \ \forall \, b \in [\etot]}
     \end{aligned}\right\}.
\end{align*}

For a value of $r_w$ that divides\footnote{If the value of $r_w$ does not divide $\etot+1$, then one can simply expand the dimension of the locator arrays in $\margarr{\ell,n}$ to the next multiple of $\etot+1$ that is greater than $r_w+(\etot+1)^2$.} $\etot+1$, the set of arrays $\margarr{\ell,n}$ that contain the locator arrays in the positions shown in Figure~\ref{fig:cons-loglog} is defined as follows.
\begin{align*}
    &\margarr{\ell,n} \\ 
     &\!\triangleq \!\left\{\bfX\!:\!
     \begin{aligned}
    &{\bfX_{[1:\etot \ell],[n-(\etot+1)^2+1:n]} = \leftmarg{\etot \ell}, } \\
    &{\bfX_{[\etot \ell+1:\etot \ell+(\etot+1)^2],[n-r_w-(\etot+1)^2+1:n]} = \topmarg{r_w+(\etot+1)^2}, } \\
    &{\bfX_{[n-(\etot+1)^2+1:n],[1:\etot \ell]} = \topmarg{\etot \ell} ,} \\
    &{\bfX_{[n-r_w-(\etot+1)^2+1:n],[\etot \ell+1:\etot \ell+(\etot+1)^2]} = \leftmarg{r_w+(\etot+1)^2} ,} \\
     \end{aligned}\right\}\! . 
\end{align*}

The set of arrays that contain the marker arrays in the positions shown in Figure~\ref{fig:cons-loglog}, is defined as follows.
\begin{align*}
\!&\markerset{\ell,n} \\
     &\!\triangleq \!\left\{\bfX\!:\!
     \begin{aligned}
     &\!{\bfX_{[\etot \ell+1:\etot \ell+(\etot+1)],[n-r_w-(\etot+1)^2-(\etot+1)+1:n-r_w-(\etot+1)^2]} } \\
    &\!{\qquad = \econstarr{1,1} ,} \\
    &\!{\bfX_{[\etot \ell+(\etot+1)^2+1:\etot \ell+(\etot+1)^2+(\etot+1)],[n-(\etot+1)+1:n]}} \\
    &\!{\qquad = \econstarr{1,2} ,} \\
    &\!{\bfX_{[n-r_w-(\etot+1)^2-(\etot+1)+1:n-r_w-(\etot+1)^2],[\etot \ell+1:\etot \ell+(\etot+1)]} } \\
    &\!{\qquad =  \econstarr{2,1} ,} \\
    &\!{\bfX_{[n-(\etot+1)+1:n],[\etot \ell+(\etot+1)^2+1:\etot \ell+(\etot+1)^+(\etot+1)]}} \\
    &\!{\qquad = \econstarr{2,2} } \\
     \end{aligned}\right\}\!.
\end{align*}

We can conclude this subsection by defining the \emph{locator set} that is the set of all arrays that have the structure required by our code to recover the indices of the inserted or deleted columns and rows. The locator set is the intersection of all the previously defined sets.

\begin{definition}[Locator Set]
We define the following set:
\begin{align*}
    \locaset \triangleq \horzedel{\ell,n} \cap \vertedel{\ell,n} \cap \margarr{\ell,n} \cap \markerset{\ell,n}.
\end{align*}
\end{definition}

The defining parameters of $\locaset$ are only $\etot$ and $n$. By fixing those, all other parameters can be obtained from the imposed constraints. Most noteworthy parameters are $w$ and $r_w$, which are functions of $n$ and $\etot$. Moreover, we point out that a good choice for the parameter $\ell$ is $\log n$, which is mainly motivated by the redundancy optimization of the construction, thoroughly discussed in Section~\ref{sec:red}. Additionally, due to the aforementioned constraints on the locator arrays, $\etot \ell$ and $r_w + (\etot+1)^2$ need to be multiples of $(\etot+1)$ and $\etot \ell \geq \left\lceil \frac{\etot}{2} \rceil\right (\etot+1)$. For an illustration of such arrays we refer to Figure~\ref{fig:cons-loglog}. Our construction works only when $\log n$ is a multiple of $\etot +1$ since we choose $\ell = \log n$.

\subsection{Construction}

Let $\mathbb{F}_q$ denote the finite field of size $q$, $\mathbb{F}_q^n$ the vector space of length $n$ over $\mathbb{F}_q$, and the $\mathbb{F}_q^{n\times n}$ the matrix space over $\mathbb{F}_q$ of dimension $n \times n$. We write $\gabi{n}{\etot} \subseteq \mathbb{F}_2^{n \times n}$ to refer to a linear\footnote{Note that such a Gabidulin code can be represented as a set of vectors in $\mathbb{F}_{2^n}^n$ as well and is linear in $\mathbb{F}_{2^n}$. For our application, it is sufficient that such a Gabidulin code is also $\mathbb{F}_2$-linear and we will always represent the codewords as binary $n \times n$ matrices.} Gabidulin code which is able to correct any pattern of $\erow$ row and $\ecol$ column erasures in an $n \times n$ array as long as $\erow + \ecol = \etot$ \cite{gabidulin2008crisscrosserasure}. This is equivalent to stating that its minimum rank distance\footnote{for the definition of the rank distance, cf. \cite{Gabidulin_TheoryOfCodes_1985}} is at least $\etot+1$. Now we are able to present our existential construction.
\begin{construction}\label{const:t-crisscross}
The code $\codedef\subseteq \sigmatn$ is the set of arrays that belong to 
\begin{align*}
    \locaset \cap \gabi{n}{\etot}.
\end{align*}
\end{construction}

\begin{theorem}\label{thm:t-crisscross-code}
The code $\codedef$ described in Construction~\ref{const:t-crisscross} is a \tcc.
\end{theorem}

A rough concept of our construction is as follows. We assume that the decoder knows whether a $\etot$-criss-cross deletion or insertion has happened from the dimension of the received array. In our codewords, we first introduce the structure $\locaset$ to locate the indices of the inserted or deleted columns and rows. With this knowledge we can introduce erasures into the missing rows and columns and convert the deletion problem into an erasure problem which can be solved by the Gabidulin code \gabi{n}{\etot} \cite{gabidulin2008crisscrosserasure}. We call this type of decoding the \emph{locate-decode strategy}. Theorem~\ref{thm:t-crisscross-code} will be proven by providing a generic decoding strategy in the next section.

%% file: tikz_cons_loglog.tex
\begin{tikzpicture}

\def\totaldist{6.5}
\def\adist{0.09*\totaldist}
\def\bdist{3*\adist}
\def\cdist{0.6*\adist}
\def\ddist{0.25*\adist}
\def\edist{3.5*\adist}

\def\markdist{2*\cdist}

\def\apattern{horizontal lines}
\def\bpattern{south west lines}
\def\cpattern{vertical lines}
\def\dpattern{north east lines}
\def\epattern{north west lines}
\def\tpattern{dots}
\def\lpattern{crosshatch dots}

\draw (0,0) rectangle (\totaldist,\totaldist);


\draw (\bdist,\totaldist-\bdist) rectangle (\totaldist-\edist-\cdist,\totaldist);

\draw (\totaldist-\edist-\cdist,\totaldist-\bdist) rectangle (\totaldist-\cdist,\totaldist);

\draw [pattern=\apattern, pattern color = lightseagreen] (\bdist, \totaldist-\bdist) rectangle (\totaldist-\cdist-\edist,\totaldist-\bdist+\adist);
\draw [pattern=\bpattern, pattern color = lightseagreen] (\bdist, \totaldist-\bdist+\adist) rectangle (\totaldist-\cdist-\edist,\totaldist-\bdist+2*\adist);
\draw [pattern=\cpattern, pattern color = lightseagreen] (\bdist, \totaldist-\bdist+2*\adist) rectangle (\totaldist-\cdist-\edist,\totaldist);

\draw [pattern=\apattern, pattern color = bostonuniversityred] (\totaldist-\cdist-\edist, \totaldist-\bdist) rectangle (\totaldist-\cdist,\totaldist-\bdist+\adist);
\draw [pattern=\bpattern, pattern color = bostonuniversityred] (\totaldist-\cdist-\edist, \totaldist-\bdist+\adist) rectangle (\totaldist-\cdist,\totaldist-\bdist+2*\adist);
\draw [pattern=\cpattern, pattern color = bostonuniversityred] (\totaldist-\cdist-\edist, \totaldist-\bdist+2*\adist) rectangle (\totaldist-\cdist,\totaldist);

\draw[pattern=\lpattern, pattern color = mediumpersianblue] (\totaldist-\cdist,\totaldist-\bdist) rectangle (\totaldist,\totaldist);

\draw[pattern=\tpattern, pattern color = darkolivegreen] (\totaldist-\edist-\cdist,\totaldist-\bdist-\cdist) rectangle (\totaldist,\totaldist-\bdist);

\draw[fill=brilliantrose] (\totaldist-\cdist-\edist-\ddist,\totaldist-\bdist-\ddist) rectangle (\totaldist-\cdist-\edist,\totaldist-\bdist);
\draw[fill=brilliantrose] (\totaldist-\ddist,\totaldist-\bdist-\cdist-\ddist) rectangle (\totaldist,\totaldist-\bdist-\cdist);

\node[circle, fill = white, inner sep = 0 pt, opacity = 0.9, text opacity = 1] at (\totaldist-0.5*\cdist, \totaldist-0.5*\bdist) {$\leftmargo^{(1)}$};
\node[circle, fill = white, inner sep = 0 pt, opacity = 0.9, text opacity = 1] at (\totaldist-0.5*\cdist-0.5*\edist, \totaldist-\bdist-0.5*\cdist) {$\topmargo^{(1)}$};

\node (m21) at ($(\totaldist-\cdist-\edist-\ddist,\totaldist-\bdist-\ddist)+(-\markdist,-1*\markdist)$) {$\left(\econstarr{2,1}\right)^{T}$};
\draw[->, gray] (m21) --  (\totaldist-\cdist-\edist-\ddist,\totaldist-\bdist-\ddist);
\node (m22) at ($(\totaldist-\ddist,\totaldist-\bdist-\cdist-\ddist)+(-\markdist,-1*\markdist)$) {$\left(\econstarr{2,2}\right)^{T}$};
\draw[->, gray] (m22) -- (\totaldist-\ddist,\totaldist-\bdist-\cdist-\ddist)  ;


\draw (0,\cdist+\edist) rectangle (\bdist,\totaldist-\bdist);

\draw (0,\cdist) rectangle (\bdist,\cdist+\edist);

\draw [pattern=\apattern, pattern color = lightseagreen] (0, \cdist+\edist) rectangle (\adist,\totaldist-\bdist);
\draw [pattern=\bpattern, pattern color = lightseagreen] (\adist, \cdist+\edist) rectangle (2*\adist,\totaldist-\bdist);
\draw [pattern=\cpattern, pattern color = lightseagreen] (2*\adist, \cdist+\edist) rectangle (\bdist,\totaldist-\bdist);

\draw [pattern=\apattern, pattern color = bostonuniversityred] (0, \cdist) rectangle (\adist,\edist+\cdist);
\draw [pattern=\bpattern, pattern color = bostonuniversityred] (\adist, \cdist) rectangle (2*\adist,\edist+\cdist);
\draw [pattern=\cpattern, pattern color = bostonuniversityred] (2*\adist, \cdist) rectangle (\bdist,\edist+\cdist);

\draw[pattern=\tpattern, pattern color = darkolivegreen] (0,0) rectangle (\bdist,\cdist);

\draw[pattern=\lpattern, pattern color = mediumpersianblue] (\bdist,0) rectangle (\bdist+\cdist,\cdist+\edist);

\draw[fill=brilliantrose] (\bdist,\cdist+\edist) rectangle (\bdist+\ddist,\cdist+\edist+\ddist);
\draw[fill=brilliantrose] (\bdist+\cdist,0) rectangle (\bdist+\cdist+\ddist,\ddist);

\node[circle, fill = white, inner sep = 0 pt, opacity = 0.9, text opacity = 1] at (\bdist+0.6*\cdist,0.5*\cdist+0.5*\edist) {$\leftmargo^{(2)}$};
\node[circle, fill = white, inner sep = 0 pt, opacity = 0.9, text opacity = 1] at (0.5*\bdist,0.5*\cdist) {$\topmargo^{(2)}$};

\node (m11) at ($(\bdist+\ddist,\cdist+\edist+\ddist)+(1.5*\markdist,0.5*\markdist)$) {$\econstarr{2,1}$};
\draw[->, gray] (m11) -- (\bdist+\ddist,\cdist+\edist+\ddist)  ;
\node (m12) at ($(\bdist+\cdist+\ddist,\ddist)+(1.5*\markdist,0.5*\markdist)$) {$\econstarr{2,2}$};
\draw[->, gray] (m12) -- (\bdist+\cdist+\ddist,\ddist)  ;


\def\xdist{0.65*\totaldist}
\def\ydist{0.25*\totaldist}
\def\rxdist{0.2*\totaldist}
\def\rydist{0.04*\totaldist}
\def\bindist{0.1*\ydist}
\def\wzdist{0.65*\xdist}
\def\rzdist{\xdist-\wzdist}

\coordinate (zero) at (1*\rxdist,\totaldist+\rydist);

\coordinate (ul) at (zero);
\coordinate (ol) at ($(zero)+(0,\ydist)$);
\coordinate (or) at ($(zero)+(\xdist,\ydist)$);;
\coordinate (ur) at ($(zero)+(\xdist,0)$);;

\draw[dashed, gray] (\bdist,\totaldist-\adist) -- (ul);
\draw[dashed, gray] (\bdist,\totaldist) -- (ol);
\draw[dashed, gray] (\totaldist-\cdist,\totaldist-\adist) -- (ur);
\draw[dashed, gray] (\totaldist-\cdist,\totaldist) -- (or);

\draw (zero) rectangle ($(zero)+(\xdist,\ydist)$);


\foreach \x/\y in {0/1,2/3,4/5,6/7,8/9} 
    {
    \draw[pattern=\epattern, pattern color = dartmouthgreen] ($(zero)+(0,\x*\bindist)$) rectangle ($(zero)+(\wzdist,\y*\bindist)$);
    \draw[pattern=\epattern, pattern color = debianred] ($(zero)+(\wzdist,\x*\bindist)$) rectangle ($(zero)+(\xdist,\y*\bindist)$);
    }
\foreach \x/\y in {1/2,3/4,5/6,7/8,9/10}
    {
    \draw[pattern=\dpattern, pattern color = dartmouthgreen] ($(zero)+(0,\x*\bindist)$) rectangle ($(zero)+(\wzdist,\y*\bindist)$);
    \draw[pattern=\dpattern, pattern color = debianred] ($(zero)+(\wzdist,\x*\bindist)$) rectangle ($(zero)+(\xdist,\y*\bindist)$);
    }

\node[text=gray] at ($(zero)+(0.5*\xdist, 1.25*\ydist)$) {\begin{varwidth}{4cm}\centering \footnotesize deletion-correcting codes with window constraint\end{varwidth}};



\def\decodist{0.005*\totaldist}

\draw [decorate,decoration={brace,amplitude=10pt,raise=4pt},yshift=0pt, color = gray]
(-\decodist,\totaldist-\bdist) -- (-\decodist,\totaldist) node [black,midway,xshift=-1.0cm, color = gray] {\footnotesize $\etot \log(n)$};

\draw [decorate,decoration={brace,amplitude=10pt,raise=4pt},yshift=0pt, color = gray]
(-\decodist,\cdist) -- (-\decodist,\cdist+\edist) node [black,midway,xshift=-0.8cm, color = gray] {\footnotesize $r_w$};

\draw [decorate,decoration={brace,amplitude=10pt,raise=4pt},yshift=0pt, color = gray]
(-\decodist,0) -- (-\decodist,\cdist) node [black,midway,xshift=-1.0cm, color = gray] {\footnotesize $(\etot+1)^2$};

\draw [decorate,decoration={brace,mirror,amplitude=10pt,raise=4pt},yshift=0pt, color = gray]
(0,-\decodist) -- (\bdist,-\decodist) node [black,midway,yshift=-0.8cm, color = gray] {\footnotesize $\etot \log(n)$};

\draw [decorate,decoration={brace,mirror,amplitude=10pt,raise=4pt},yshift=0pt, color = gray]
(\bdist,-\decodist) -- (\bdist+\cdist,-\decodist) node [black,midway,yshift=-0.8cm, xshift=0.2cm, color = gray] {\footnotesize $(\etot+1)^2$};

\draw [decorate,decoration={brace,amplitude=10pt,raise=4pt},yshift=0pt, color = gray]
($(ul)+(-\decodist,0)$) -- ($(ol)+(-\decodist,0)$) node [black,midway,xshift=-1.0cm, color = gray] {\footnotesize $\log (n)$};

\end{tikzpicture}

%% file: decoder_loglog.tex
Assume a codeword $\bfC \in \codedef$ is transmitted and let $\ecol$ and $\erow$ be such that $\ecol + \erow =t$. 

The decoder receives an array $ \widetilde{\bfC} \in \Sigma ^{(n-\erow)\times(n-\ecol)}$ obtained from $\bfC$ by $\erow$ row and $\ecol$ column deletions or an array $ \widetilde{\bfC} \in \Sigma ^{(n+\erow)\times(n+\ecol)}$ obtained from $\bfC$ by $\erow$ row and $\ecol$ column insertions. The dimension of the received array is assumed to be known to the decoder. As mentioned before we first focus on locating the indices of inserted/deleted rows and columns.  

\subsection{Locating the indices}
 Let us denote the set of indices of the rows and columns that got inserted or deleted by $\rdset \subset [n+\erow]$ and $\cdset \subset [n+\ecol]$, respectively, with $|\rdset|+|\cdset|=\erow + \ecol = \etot$. For clarity of presentation, we first present the decoding strategy for locating deletions. Subsequently, we present the decoding strategy for locating insertions.

\begin{claim} \label{cl:marker-locating-deletion}
Given the array $\widetilde{\bfC}$ affected by $\erow$ row and $\ecol$ column deletions, the marker arrays $\tileconstarr{1,1}$, $\tileconstarr{1,2}$, $\tileconstarr{2,1}$, and $\tileconstarr{2,2}$ can be located.
\end{claim}
\begin{IEEEproof}
We will focus on locating the arrays $\tileconstarr{2,1}$ and $\tileconstarr{2,2}$. 
A similar proof can be given to find the arrays $\tileconstarr{1,1}$ and $\tileconstarr{1,2}$ by exploiting the symmetric properties of $\topmargo^{(1)}$ and $\leftmargo^{(1)}$.

Using Claim~\ref{cl:leftmargo-deletions} we can locate the leftmost column of $\widetilde{\leftmargo}^{(2)}$. Moreover, the bottom row of $\widetilde{\leftmargo}^{(2)}$ is given directly by the position of $\leftmargo^{(2)}$ in the codeword itself. 

Note that we only have to detect row or column deletions rather than exactly locating them. 
Our goal is to locate the rightmost column of $\widetilde{\leftmargo}^{(2)}$. Recall that in the last $\etot +1$ rows of $\widetilde{\leftmargo}^{(2)}$ there is for sure an $\widetilde{\leftmargo'}$ which originates from $\leftmargo'$ affected by possible row and column deletions. By a similar argument as in Claim~\ref{cl:leftmargo-deletions}, one can detect the number of column deletions that happened in each subarray of $\leftm{\prime}$ of $\etot +1 $ consecutive columns, starting from the left. 
This stems from the fact that number of column runs in $\widetilde{\leftmargo}^{\prime}$ is $\etot +1$ minus the number of row deletions in $\leftm{\prime}$. Recall that from Claim~\ref{cl:leftmargo-deletions} we know the number of row deletions in $\leftm{\prime}$. The remaining ingredient is to know when $\widetilde{\leftmargo}^\prime$ ends. This is guaranteed since the columns of the marker $\econstarr{2,2}$ are the complement of the last $\etot +1$ columns in $\leftm{\prime}$. Therefore, we are guaranteed to have at least one column of $\tileconstarr{2,2}$ marking the end of $\widetilde{\leftmargo}^\prime$ even in the presence of row or column deletions. 

Given the rightmost column of $\widetilde{\leftmargo}^{(2)}$ we focus on locating the marker $\tileconstarr{2,1}$ and therefore the topmost row of $\widetilde{\leftmargo}^{(2)}$. Recall that $\leftmargo^{(2)}$ consists of $\frac{s}{\etot+1}$ arrays $\leftm{\prime}$ stacked on top of each other. Using the same argument as in Claim~\ref{cl:leftmargo-deletions}, we can locate every row deletion in $\leftmargo^{(2)}$ until the topmost subarray $\leftm{\prime}$. This is true even in the presence of column deletions, since column deletions do not change the fact that every $\etot + 1$ consecutive rows in $\leftmargo^{(2)}$ are different. By the choice of the marker $\econstarr{2,1}$, we ensure that the number of row deletions in the topmost $\leftm{\prime}$ can be detected. This is true because the marker has the same structure of the first $\etot+1$ columns of $\leftm{\prime}$.
\end{IEEEproof}

\begin{lem}\label{lem:index-recover-manab-del}
Given the array $\widetilde{\bfC}$ affected by $\erow$ row and $\ecol$ column deletions, any row with index $i \in \rdset$ such that $1 \leq i \leq \etot \ell$ or $n-r_w-(\etot+1)^2 < i \leq n$ can be exactly recovered.

Similarly, any column index $j \in \cdset$ such that $1 \leq j \leq \etot \ell$ or $n-r_w-(\etot+1)^2 < j \leq n$ can be exactly recovered.
\end{lem}
\begin{IEEEproof}
We focus on recovering the indices of the deleted rows such that $i \in \rdset$ and the indices of the deleted columns such that $j \in \cdset$ that satisfy $1 \leq i \leq \etot \ell$ and $n-r_w-(\etot+1)^2 < j \leq n$. Recovering the remaining indices of the statement follows by the symmetry of the construction. 

From Claim~\ref{cl:marker-locating-deletion}, the location of $\widetilde{\leftmargo}^{(1)}$ and $\widetilde{\topmargo}^{(1)}$ in $\widetilde{\bfC}$ can be exactly recovered. Therefore, by Claim~\ref{cl:leftmargo-deletions} we can locate any column deletions with indices $n-r_w-(\etot+1)^2 < j \leq n$ by decoding $\widetilde{\topmargo}^{(1)}$. Consequently, having the location of $\widetilde{\leftmargo}^{(1)}$ and using Claim~\ref{cl:leftmargo-deletions}, we can recover the indices of the deleted rows that satisfy $1 \leq i \leq \etot \ell$. 

Similarly, we can obtain the indices with $n-r_w-(\etot+1)^2 < i \leq n$ and $1 \leq j \leq \etot \ell$.
\end{IEEEproof}

\begin{lem}\label{lem:index-recover-delcode-del}
Given the array $\widetilde{\bfC}$ affected by $\erow$ row and $\ecol$ column deletions, any row index $i \in \rdset$ such that $\etot \ell < i \leq n-r_w-(\etot+1)^2$  can be exactly recovered.

Similarly, any column index $j \in \cdset$ such that $\etot \ell < j \leq n-r_w-(\etot+1)^2$ can be exactly recovered.
\end{lem}
\begin{IEEEproof}
We start by proving that the column indices can be recovered. We want to leverage the structure imposed by the set $\horzedel{\ell,n}$. For an array $\bfC\in\horzedel{\ell,n}$, each row of the subarray $\bfC_{[1:\etot\ell],[\etot\ell+1:n-(\etot+1)^2]}$ 
is encoded using a binary systematic $\etot$-\insdel-correcting code. In addition, the columns $\bfC_{[1:\etot\ell],j}$ such that $\etot \ell <j\leq n-r_w-(\etot+1)^2$ are the systematic part of this code. Recall that the rows are divided into $\etot$ blocks, each of size $\ell$, where in each block the columns $\etot <j\leq n-r_w-(\etot+1)^2$ satisfy the window constraint. 
We assume that at least one column in this interval is deleted. Therefore, at most $(\etot-1)$ rows can be deleted in $\bfC$. This means that there exists at least one block of $\ell$ rows that is not affected by any row deletion. For the deletion case, by Lemma~\ref{lem:index-recover-manab-del} we can locate this block. By Claim~\ref{cl:deletion-locate-set} we can recover the indices of the columns deleted within the range $\etot <j\leq n-r_w-(\etot+1)^2$. 

Similarly, we can obtain the indices with $\etot \ell < i \leq n-r_w-(\etot+1)^2$ by leveraging the structure imposed by $\vertedel{\ell,n}$ using Claim~\ref{cl:deletion-locate-set}.
\end{IEEEproof}

We now present the equivalent results for the insertion case. In the following statements we only highlight the differences from the deletion case. The main difference is that we need to tackle possible insertion patterns which can create block confusions in our received array $\widetilde{\bfC}$. In general, the strategy is as follows. Since all rows and columns of the codeword $\bfC$ satisfy the window constraint, even after $\etot$-criss-cross insertions, we can exploit the fact that insertions can only create a confusion of length at most $2\etot$ with at most $\etot$ rows/columns of the original array. Therefore, we can delete the block confusions and turn the insertion locating problem into a deletion locating problem, which we have shown earlier how to solve.

\begin{claim} \label{cl:marker-locating-insertion}
Given the array $\widetilde{\bfC}$ affected by $\erow$ row and $\ecol$ column insertions, the marker array $\tileconstarr{1,1}$ can be located up to row and column block confusion of length at most $2\etot$, and $\tileconstarr{1,2}$ can be located up to column block confusion of length at most $2\etot$ and row block confusion of length at most $2\etot+1$. The marker array $\tileconstarr{2,1}$ can be located up to row and column block confusion of length at most $2\etot$, and $\tileconstarr{1,2}$ can be located up to column block confusion of length at most $2\etot$ and row block confusion of length at most $2\etot+1$. 
\end{claim}
\begin{IEEEproof}
We have the same preliminary information as in Claim~\ref{cl:leftmargo-insertions} and \ref{cl:marker-locating-deletion}. However, if the inserted rows/columns create block confusions exactly at the border of the arrays $\leftmargo^{(1)}$ and $\topmargo^{(1)}$, or  $\leftmargo^{(2)}$ and $\topmargo^{(2)}$ we can locate the desired arrays only up to the length of the block confusion. By construction, the length of the block confusion is limited to the parameter range given in the statement of the claim.
Moreover, we use the following strategy for locating the border between $\leftmargo^{(1)}$ and $\topmargo^{(1)}$ in presence of block confusions, where the same can be applied to the border of $\leftmargo^{(2)}$ and $\topmargo^{(2)}$ by symmetry of the construction. The idea is to use a block of \insdel\ codes to resolve the row block confusion. Note that there must be at least one \insdel\ code block which is not affected by insertions due to the existence of $\etot$ \insdel\ code subarrays. Moreover, we can determine the affected arrays by Claim~\ref{cl:leftmargo-insertions}. We declare the border of the \insdel\ code to be the first expected row index of $\topmargo^{(1)}$. Note that by declaring a wrong border, we add at most $\etot$ insertions to the \insdel\ code. By Claim~\ref{cl:insertion-locate-set} we can decode the \insdel\ code. The correct border can then be determined by the first index which is not marked as an insertion in the \insdel\ code subarray. In case no insertions are declared at the beginning of the subarray, then the chosen border is correct.

Moreover, in case $\tileconstarr{1,2}$ and $\tileconstarr{2,2}$ are located up to a column block confusion and a row block confusion, respectively, one can simply ignore the confusion to locate $\tileconstarr{1,1}$ and $\tileconstarr{2,1}$ since the inserted columns must follow the structure of $\topmargo^{(1)}$ and $\leftmargo^{(2)}$ to create a confusion.
\end{IEEEproof}

Observe that the marker arrays $\tileconstarr{1,2}$ and $\tileconstarr{2,2}$ can be located up to a row, or respectively a column block, confusion of length at most $2\etot+1$. The confusion may contain more than $\etot$ original rows/columns of the original array. However, since the indices of the block confusions are within the index range of the \insdel\ codes, we can use those as stated in Lemma~\ref{lem:index-recover-delcode-insertion} to tackle this problem.

\begin{lem}\label{lem:index-recover-manab-insertion}
Given the array $\widetilde{\bfC}$ affected by $\erow$ row and $\ecol$ column insertions, any row with index $i \in \rdset$ such that $1 \leq i \leq \etot \ell$ or $n-r_w-(\etot+1)^2 < i \leq n$  can be recovered up to a row block confusion of length at most $2\etot$ consisting of at at most $\etot$ rows of the original array $\bfC$.
Similarly, any column index $j \in \cdset$ such that $1 \leq j \leq \etot \ell$ or $n-r_w-(\etot+1)^2 < j \leq n$ can be recovered  up to a column block confusion of length at most $2\etot$ consisting of at at most $\etot$ columns of the original array $\bfC$.
\end{lem}
\begin{IEEEproof}
We focus on recovering the indices of the inserted rows such that $i \in \rdset$ and the indices of the inserted columns such that $j \in \cdset$ that satisfy $1 \leq i \leq \etot \ell$ and $n-r_w-(\etot+1)^2 < j \leq n$. Recovering the remaining indices of the statement follows by symmetry of the construction. In general, we can use the same strategy as presented for the deletion case using Claim~\ref{cl:leftmargo-insertions} and~\ref{cl:marker-locating-insertion}. 
In case the marker arrays are located up to a confusion, one can ignore the rows/columns which created a confusion, \ie we only consider one collection of the rows/columns of the confusion, since the inserted rows/columns must satisfy the fixed structure of the locator arrays $\topmargo^{(1)}$, $\leftmargo^{(1)}$, $\topmargo^{(2)}$, and $\leftmargo^{(2)}$.
\end{IEEEproof}

\begin{lem}\label{lem:index-recover-delcode-insertion}
Given the array $\widetilde{\bfC}$ affected by $\erow$ row and $\ecol$ column insertions, any row index $i \in \rdset$ such that $\etot \ell < i \leq n-r_w-(\etot+1)^2$ can be recovered up to a row block confusion of length at most $2\etot$ consisting of at at most $\etot$ rows of the original array $\bfC$.
Similarly, any column index $j \in \cdset$ such that $\etot \ell < j \leq n-r_w-(\etot+1)^2$ can be recovered up to a column block confusion of length at most $2\etot$ consisting of at most $\etot$ columns of the original array $\bfC$.
\end{lem}
\begin{IEEEproof}
We start by proving that the column indices can be recovered. We want to leverage the structure imposed by the set $\horzedel{\ell,n}$. We assume that at least one column in this interval is inserted. This means that there exists at least one block of $\ell$ rows that is not affected by any row insertion. By Lemma~\ref{lem:index-recover-manab-insertion} we can locate this block by remarking that at least one insertion is needed to create a block confusion and therefore affect the block. By Claim~\ref{cl:insertion-locate-set} we can recover the indices of the columns inserted within the range $\etot <j\leq n-r_w-(\etot+1)^2$ up to a block confusion of length at most $2\etot$ containing at most $\etot$ columns of the original array.
Similarly, we can obtain the indices with $\etot \ell < i \leq n-r_w-(\etot+1)^2$ by leveraging the structure imposed by $\vertedel{\ell,n}$, cf., Lemma~\ref{lem:index-recover-manab-insertion}, and Claim~\ref{cl:insertion-locate-set}.
\end{IEEEproof}

\subsection{Recovering the transmitted array}
Now we can present the full proof of our code construction.
\begin{IEEEproof}[Proof of Theorem~\ref{thm:t-crisscross-code}]
If a $\etot$-criss-cross deletion happened, we can apply Lemma~\ref{lem:index-recover-manab-del}~and~\ref{lem:index-recover-delcode-del} to determine the sets of indices $\rdset$ and $\cdset$. Then, for all $i \in \rdset$ and $j \in \cdset$ the decoder inserts row or column erasures in $\widetilde{\bfC}$ starting from the smallest index. Now the decoder applies a Gabidulin criss-cross erasure decoder to determine the values of the erased symbols \cite{gabidulin2008crisscrosserasure}. 

In case of a $\etot$-criss-cross insertion we apply Lemma~\ref{lem:index-recover-manab-insertion}~and~\ref{lem:index-recover-delcode-insertion}. The decoder deletes the inserted rows/columns which their positions are exactly recovered. For each block confusion, the decoder deletes the whole block confusion. This deletion strategy deletes at most $\etot$ rows/columns of the original array, since all rows/columns follow the window constraint. Thus, the decoder inserts row/column erasures in $\widetilde{\bfC}$ starting from the smallest index and applies a Gabidulin criss-cross erasure decoder to determine the values of the erased symbols \cite{gabidulin2008crisscrosserasure}. 
\end{IEEEproof}

%% file: redundancy_loglog.tex
In this section we perform an analysis of the redundancy of our code denoted by $\red$. We will refer to the redundancy of each individual set $\gabi{\etot}{n}$, $\locaset$, $\horzedel{\ell,n}$, $\vertedel{\ell,n}$, $\margarr{\ell,n}$, $\winarr{\ell,w}$ and $\markerset{\ell,n}$ by $R_{\ast}(n,\etot)$, where $\ast$ is replaced with the corresponding set letter.
In the following, we give an intuition behind the computations of the redundancy.

Since $\codedef = \locaset \cap \gabi{\etot}{n}$ and due to the fact that the Gabidulin code is a linear code, we can compute the code redundancy as follows.
\begin{equation*}
    \red = \redloca + \redgab
\end{equation*}

Moreover, since the intersected sets in the locator set $\locaset$ impose constraints on disjoint positions in the $n \times n$ arrays, we can further split the redundancy as follows
\begin{align*}
    \redloca = \redho + \redve + \redmanab + \redmarker .
\end{align*}
The sets $\horzedel{\ell,n}$ and $\vertedel{\ell,n}$ impose similar constraints: $\etot$ disjoint subarrays constrained with the \wcons\ where each row is protected by a systematic $\etot$-deletion correcting code from \cite{sima2020systematictdel}.
\begin{claim}\label{cl:red-hovedel-comb}
The redundancy resulting from the constraints imposed by the two sets $\horzedel{\ell,n}$ and $\vertedel{\ell,n}$ is bounded as
\begin{align*}
     \redho + \redve \leq 2\etot ( \redw{w} + \log (n) \cdot r_w ),
\end{align*}
where $w = n - \etot \log(n) - r_w - (\etot+1)^2$ and $r_w \leq 4 \etot \log(n) + o(\log n)$.
\end{claim}
\begin{IEEEproof}
The sets $\horzedel{\ell,n}$ and $\vertedel{\ell,n}$ impose the same constraints, \ie each array belonging to any of these sets has $\etot$ subarrays protected by deletion-correcting codes with window constraints. Therefore, we have
\begin{align*}
    \redho + \redve = 2\etot ( \redw{w} + \log (n) \cdot r_w ),
\end{align*}
where $r_w$ is the length of the redundancy vector used to protect a vector of length%
\begin{equation}\label{eq:w}
    w = n - \etot \log (n) - r_w - (\etot+1)^2    
\end{equation}
against $\etot$ deletions, and $\log (n)$ is the number of protected vectors in each subarray. Recall that for any integer $\k $, the redundancy for protecting a vector of length $\k$ is bounded by $r_{\k} \leq 4\etot \log (\k)+ o(\log(\k))$~\cite{sima2020systematictdel}. 
\begin{equation}\label{eq:rw}
    r_w \leq (4 \etot+1) \log(n).
\end{equation}

We now focus on computing $\redw{w}$. To compute an upper bound on the redundancy imposed by the window constraint $\winarr{\ell,w}$ we require a lower bound on $w$. Note that using a lower bound on $w$ only increases the redundancy imposed by $\winarr{\ell,w}$. This will be clear from the following calculations. From~\eqref{eq:w} and~\eqref{eq:rw} we obtain%
\begin{equation*}
    w \geq n - (5\etot+1) \log (n) - (\etot+1)^2.
\end{equation*}
We calculate a lower bound on $\vert \winarr{\ell,w} \vert$. On a high level, our calculations are interpreted as going through each column of an array in $\winarr{\ell,w}$ and counting the number of choices for this specific column. The first column is arbitrary, thus has $2^\ell$ choices. The second column is not allowed to be the same as the one before, thus it has $(2^\ell-1)$ choices. The third column has $(2^\ell-2)$ choices, since it cannot be the same as the two preceding columns. This process continues until we reach the $(\etot+2)^\text{nd}$ column. The number of choices for this vector is $(2^\ell-(\etot+1))$. Since the restriction is imposed on an interval of $(\etot+1)$ vectors, each remaining columns has $(2^\ell-(\etot+1))$ choices. Thus, for the window constraint the following holds. 
\begin{align*}
    \lvert \winarr{\ell,w} \rvert &\geq 2^\ell \cdot (2^\ell - 1) \cdot \ldots  \cdot (2^\ell - (\etot+1)) 
    \\ &\qquad \cdot (2^\ell - (\etot+1))^{w-\etot-2} \\
    &\geq (2^\ell - (\etot+1)) ^{w} \\
    &= 2^{\ell w}\left( 1 - \frac{\etot+1}{2^\ell} \right)^{w}.
\end{align*}
We denote the redundancy resulting from the constraints imposed by the window constraint as $\redw{n-\etot \ell - r_w}$. We continue the calculations recalling that $\ell = \log(n)$.
\begin{flalign*}
    &\redw{n-\etot \ell - r_w} \\
    &\leq \ell (n-\etot \ell - r_w) - \log \left(\lvert \winarr{\ell,n-\etot \ell - r_w} \rvert \right) \\
    &\leq \log \left(\left( 1 - \frac{\etot+1}{2^\ell} \right)^{n-\etot \ell - r_w} \right) \\
    &\leq \log \left(\left( 1 - \frac{\etot+1}{n} \right)^{n} \right) - \log \left(\left( 1 - \frac{\etot+1}{n} \right)^{(5 \etot+1) \log(n) } \right) \\
    &\overset{(a)}{\leq} \log (e^{(\etot+1)}) - (5\etot+1)  \log(n) \cdot \log \left(\left( 1 - \frac{\etot+1}{n} \right) \right) \\
    &\overset{(b)}{\leq} (\etot+1) \log(e) + (5\etot+1) \log(n) \\
    &\leq (5\etot+1)  \log(n) + 2 (\etot + 1).
\end{flalign*}
We used in $(a)$ the inequality $\left( 1 - \frac{x}{n} \right) ^n \leq e^x$ and exploited in $(b)$ the fact that $\frac{1}{2} \leq (1-\frac{\etot+1}{n}) \leq 1$ for our choice of parameters and for sufficiently large $n$.

Recall that any array in $\edelarr{1}(\ell,n-\etot \ell - (\etot+1)^2)$ consists of $\log(n)$ binary $\etot$-deletion correcting codes. Therefore, we have
\begin{align*}
    &\redho  \\
    &\leq \etot \cdot \left( \underbrace{ (5\etot+1) \log(n) + 2(\etot+1) }_{\text{window constraint}} \right. \\
    &\qquad \qquad \qquad \qquad \left. + \underbrace{\log(n) \cdot (4 \etot + 1) \log(n)}_{\text{binary deletion correcting codes}} \right) \\
    &= (4 \etot^2 + \etot ) \log^2(n) + (5\etot^2+\etot)  \log(n) + 2\etot(\etot+1)^2.
\end{align*}
Since the arrays in $\vertedel{\ell,n}$ have a similar structure imposed (only transposed) and the regions of the imposed constraints are disjoint, one can conclude that 
\begin{align*}
    &\redho + \redve \\
    &\leq 2(4 \etot^2 + \etot ) \log^2(n) + 2(5\etot^2+\etot)  \log(n) + 4\etot(\etot+1)^2.
\end{align*}

\end{IEEEproof}
Observe that the constraints for the remaining sets fix values for certain subarray boundaries. Therefore, the following can be obtained.

\begin{claim}\label{claim:red-locaset}
The redundancy $\redloca$ resulting from the constraints imposed by the set $\locaset$ is bounded as
\begin{align*}
      \redloca &\leq (8\etot^2+2\etot) \log^2(n) + o(\log^2(n)).
 \end{align*}
\end{claim}

\begin{IEEEproof}
We argued that since the different constraints are imposed on disjoint subarrays in $\locaset$, then the redundancy $\redloca$ can be written as
\begin{align*}
    \redloca = \redho + \redve + \redmanab + \redmarker
\end{align*}
The redundancy imposed by the locator arrays and marker arrays is equal to the dimension of the subarrays with fixed entries. We can then write
\begin{align*}
     \redmanab &\leq (6\etot^3+13\etot^2+8\etot+1) \log(n), \\
     \redmarker &= 4(\etot+1)^2.
 \end{align*}
The other terms of the redundancy in $\redloca$ are computed in Claim~\ref{cl:red-hovedel-comb}.
\end{IEEEproof}

We can conclude this section with the statement on the redundancy $\red$ of the code $\codedef$ presented in Construction~\ref{const:t-crisscross}. Note that the redundancy added by the Gabidulin code is $\etot n$.
\begin{lem}
The redundancy of the code $\codedef$ is bounded as
\begin{align*}
    \red \leq \etot n + (8\etot^2+2\etot) \log^2(n) + o(\log^2(n)).
\end{align*}
\end{lem}
\begin{IEEEproof}
By construction we have that $\codedef = \locaset \cap \gabi{n}{\etot}$. By Claim~\ref{claim:red-locaset} we have that
\begin{align*}
    \lvert \locaset \rvert \geq \frac{2^{n^2}}{n^{\left((8\etot^2+2\etot)+o(1) \right) \log(n)}}.
\end{align*}
From \cite{gabidulin2008crisscrosserasure} we have that $\lvert \gabi{n}{\etot} \rvert = \frac{2^{n^2}}{2^{\etot n}}$. Further, due to the fact that $\gabi{n}{\etot}$ is a linear code, there exists a coset such that the following is satisfied by means of the pigeon hole principle.
\begin{align*}
    \vert \codedef \vert &\geq 2^{n^2} \cdot \underbrace{\frac{1}{2^{\etot n}}}_{\textrm{Gabidulin Code}} \cdot \underbrace{\frac{1}{n^{\left((8\etot^2+2\etot)+o(1) \right) \log(n)}}}_{\textrm{Locator Set}}
\end{align*}
Hence, we can conclude that the total redundancy of the $\codedef$ satisfies
\begin{align*}
    \red &= n^2 - \log( \lvert \codedef \rvert ) \\
    &\leq \etot n + (8\etot^2+2\etot) \log^2(n) + o(\log^2(n)).
\end{align*}%
\end{IEEEproof}

%% file: conclusion_loglog.tex
In this work we have considered the $\etot$-criss-cross insertion/deletion problem in binary arrays. First, we have shown that the one-dimensional insertion-deletion equivalence also holds in the two-dimensional array setting. Moreover, we have shown that the asymptotic lower bound on the redundancy for any $\etot$-criss-cross correcting code  is $\redopt \geq \etot n + \etot \log (n) - \mathcal{O}(1)$. We have presented our \tcc\ construction which is based on the strategy of transforming the insertion/deletion problem to an erasure problem. The redundancy of the constructed  \tcc\ is $\mathcal{O}(\etot^2 \log^2 (n))$ far from the derived lower bound. We note that given an order optimal systemtiac construction of $n$-ary $\etot$-{\insdel}-correcting codes, we could improve our construction such that the redundancy is only $\mathcal{O}(\etot^3 \log(n))$ far from the derived lower bound. This results from replacing the $2 \etot \log n$ binary $\etot$-{\insdel}-correcting codes indexing the columns/rows by $2 \etot$ $n$-ary $\etot$-{\insdel}-correcting codes. 

On a final note, improvements on the problem of coding for \insdel\ errors in arrays remain possible. It would be interesting to generalize the problem to study possible combinations of simultaneous insertions and deletions in arrays. In this case, generalizing the one-dimensional equivalence between insertions and deletions amounts to generalizing the equivalence between criss-cross insertion correcting codes and criss-cross deletion correcting codes (Theorem~\ref{theorem:equiv}) to proving that a code able to correct $\etot$ criss-cross deletions can correct any number of $\tr$ row insertions (or deletions) and $\tc$ column deletions (or insertions) such that $\etot =\tr+\tc$. The next step would then be finding a code construction, with redundancy close to the lower bound derived in this paper, that can correct a mixtures of \insdel\  column and row errors. Further research topics in this direction include studying the characteristics of the \insdel\ spheres of an array $\mathbf{X}$.

\section*{Acknowledgements}
We thank the associate editor and the anonymous reviewers for their valuable comments that contributed to the improvement of the quality of this work. Further, we want to thank Evagoras Stylianou for his observations concerning the decoder for detecting insertions, leading to a better explanation of the decoder.

%% file: long.bbl
\begin{thebibliography}{10}
\providecommand{\url}[1]{#1}
\csname url@samestyle\endcsname
\providecommand{\newblock}{\relax}
\providecommand{\bibinfo}[2]{#2}
\providecommand{\BIBentrySTDinterwordspacing}{\spaceskip=0pt\relax}
\providecommand{\BIBentryALTinterwordstretchfactor}{4}
\providecommand{\BIBentryALTinterwordspacing}{\spaceskip=\fontdimen2\font plus
\BIBentryALTinterwordstretchfactor\fontdimen3\font minus
  \fontdimen4\font\relax}
\providecommand{\BIBforeignlanguage}[2]{{%
\expandafter\ifx\csname l@#1\endcsname\relax
\typeout{** WARNING: IEEEtran.bst: No hyphenation pattern has been}%
\typeout{** loaded for the language `#1'. Using the pattern for}%
\typeout{** the default language instead.}%
\else
\language=\csname l@#1\endcsname
\fi
#2}}
\providecommand{\BIBdecl}{\relax}
\BIBdecl

\bibitem{ISITcriss}
L.~Welter, R.~Bitar, A.~Wachter-Zeh, and E.~Yaakobi, ``Multiple criss-cross
  deletion correcting codes,'' \emph{IEEE International Symposium on
  Information Theory (ISIT)}, 2021.

\bibitem{Heckel_A-Characterization-of-the-DNA-Storage-Channel_19}
\BIBentryALTinterwordspacing
R.~Heckel, G.~Mikutis, and R.~N. Grass, ``{A Characterization of the DNA Data
  Storage Channel},'' \emph{Scientific Reports}, vol.~9, no.~1, p. 9663, 2019.
  [Online]. Available: \url{https://doi.org/10.1038/s41598-019-45832-6}
\BIBentrySTDinterwordspacing

\bibitem{SalaSchoeny-Sync_TransComm}
F.~{Sala}, C.~{Schoeny}, N.~{Bitouzé}, and L.~{Dolecek}, ``Synchronizing files
  from a large number of insertions and deletions,'' \emph{IEEE Transactions on
  Communications}, vol.~64, no.~6, pp. 2258--2273, June 2016.

\bibitem{venkataramanan2010interactive}
R.~Venkataramanan, H.~Zhang, and K.~Ramchandran, ``Interactive low-complexity
  codes for synchronization from deletions and insertions,'' in \emph{48th
  Annual Allerton Conference on Communication, Control, and Computing
  (Allerton)}, 2010, pp. 1412--1419.

\bibitem{7185447}
R.~{Venkataramanan}, V.~{Narasimha Swamy}, and K.~{Ramchandran},
  ``Low-complexity interactive algorithms for synchronization from deletions,
  insertions, and substitutions,'' \emph{IEEE Transactions on Information
  Theory}, vol.~61, no.~10, pp. 5670--5689, 2015.

\bibitem{yazdi2013deterministic}
S.~S.~T. Yazdi and L.~Dolecek, ``A deterministic polynomial-time protocol for
  synchronizing from deletions,'' \emph{IEEE Transactions on Information
  Theory}, vol.~60, no.~1, pp. 397--409, 2013.

\bibitem{ma2011efficient}
N.~{Ma}, K.~{Ramchandran}, and D.~{Tse}, ``Efficient file synchronization: A
  distributed source coding approach,'' in \emph{IEEE International Symposium
  on Information Theory Proceedings}, 2011, pp. 583--587.

\bibitem{DolecekAnan_Sync_2007}
L.~{Dolecek} and V.~{Anantharam}, ``Using {Reed–Muller} {RM} (1, m) codes
  over channels with synchronization and substitution errors,'' \emph{IEEE
  Transactions on Information Theory}, vol.~53, no.~4, pp. 1430--1443, April
  2007.

\bibitem{Levenshtein-binarycodesCorrectingDeletions}
{V.I. Levenshtein}, ``{Binary codes capable of correcting deletions, insertions
  and reversals (in Russian)},'' \emph{Doklady Akademii Nauk SSR}, vol. 163,
  no.~4, pp. 845--848, 1965.

\bibitem{VarshTene-SingleDeletion1965}
R.~R. Varshamov and G.~M. Tenengolts, ``{Codes which correct single asymmetric
  errors (in Russian)},'' \emph{Automatika i Telemkhanika}, vol. 161, no.~3,
  pp. 288--292, 1965.

\bibitem{GuruswamiWang-HighNoiseHighRateDeletions_2017}
V.~Guruswami and C.~Wang, ``Deletion codes in the high-noise and high-rate
  regimes,'' \emph{IEEE Transactions on Information Theory}, vol.~63, no.~4,
  pp. 1961--1970, Apr. 2017.

\bibitem{brakensiek2017efficient}
J.~Brakensiek, V.~Guruswami, and S.~Zbarsky, ``Efficient low-redundancy codes
  for correcting multiple deletions,'' \emph{IEEE Transactions on Information
  Theory}, vol.~64, no.~5, pp. 3403--3410, 2017.

\bibitem{hanna2018guess}
S.~K. Hanna and S.~El~Rouayheb, ``Guess \& check codes for deletions,
  insertions, and synchronization,'' \emph{IEEE Transactions on Information
  Theory}, vol.~65, no.~1, pp. 3--15, 2018.

\bibitem{Gabrys-TwoDeletions_2018}
R.~{Gabrys} and F.~{Sala}, ``Codes correcting two deletions,'' \emph{IEEE
  Transactions on Information Theory}, vol.~65, no.~2, pp. 965--974, Feb 2019.

\bibitem{Sima-TwoDeletions_2019}
J.~{Sima}, N.~{Raviv}, and J.~{Bruck}, ``Two deletion correcting codes from
  indicator vectors,'' \emph{IEEE Transactions on Information Theory}, pp.
  1--1, 2019.

\bibitem{SimaBruck-kDeletions_2020}
J.~{Sima} and J.~{Bruck}, ``On optimal k-deletion correcting codes,''
  \emph{IEEE Transactions on Information Theory}, pp. 1--1, 2020.

\bibitem{guruswami2020explicit}
V.~Guruswami and J.~H{\r a}stad, ``Explicit two-deletion codes with redundancy
  matching the existential bound,'' in \emph{Proceedings of the 2021 ACM-SIAM
  Symposium on Discrete Algorithms (SODA)}.\hskip 1em plus 0.5em minus
  0.4em\relax SIAM, 2021, pp. 21--32.

\bibitem{sima2020systematictdel}
J.~{Sima}, R.~{Gabrys}, and J.~{Bruck}, ``Optimal systematic $t$-deletion
  correcting codes,'' in \emph{IEEE International Symposium on Information
  Theory (ISIT)}, 2020, pp. 769--774.

\bibitem{krishnamurthy2019trace}
A.~Krishnamurthy, A.~Mazumdar, A.~McGregor, and S.~Pal, ``Trace reconstruction:
  Generalized and parameterized,'' \emph{arXiv preprint arXiv:1904.09618},
  2019.

\bibitem{SchoenyWachterzehGabrysYaakobi-BurstDeletions-journal}
C.~Schoeny, A.~{Wachter-Zeh}, R.~Gabrys, and E.~Yaakobi, ``Codes correcting a
  burst of deletions or insertions,'' \emph{IEEE Transactions on Information
  Theory}, vol.~63, no.~4, pp. 1971--1985, Apr. 2017.

\bibitem{smith2017interleaved}
D.~Smith, T.~G. Swart, K.~A. Abdel-Ghaffar, H.~C. Ferreira, and L.~Cheng,
  ``Interleaved constrained codes with markers correcting bursts of insertions
  or deletions,'' \emph{IEEE Communications Letters}, vol.~21, no.~4, pp.
  702--705, 2017.

\bibitem{bakirtas2021database}
S.~Bakirtas and E.~Erkip, ``Database matching under column deletions,''
  \emph{arXiv preprint arXiv:2105.09616}, 2021.

\bibitem{roth1991maximum}
R.~M. Roth, ``Maximum-rank array codes and their application to crisscross
  error correction,'' \emph{IEEE Transactions on Information Theory}, vol.~37,
  no.~2, pp. 328--336, 1991.

\bibitem{gabidulin2008crisscrosserasure}
E.~M. Gabidulin and N.~I. Pilipchuk, ``Error and erasure correcting algorithms
  for rank codes,'' \emph{Designs, Codes and Cryptography}, vol.~49, pp.
  105--122, 2008.

\bibitem{LundGabidulinHonary-NewFamilyOptimalCorrectingTermRankErrors_2000}
D.~Lund, E.~M. Gabidulin, and B.~Honary, ``{A new family of optimal codes
  correcting term rank errors},'' in \emph{IEEE International Symposium on
  Information Theory Proceedings}, June 2000, p. 115.

\bibitem{Sidorenko-ClassCorrectingErrorsLatticeConfiguration_1976}
V.~R. Sidorenko, ``Class of correcting codes for errors with a lattice
  configuration,'' \emph{Problemy Reredachi Informatsii}, vol.~12, no.~3, pp.
  165--171, Mar. 1976.

\bibitem{BlaumBruck-ArrayCodesCorrectionCrisscrossErrors_IEEE-IT2000}
M.~Blaum and J.~Bruck, ``{MDS} array codes for correcting a single criss-cross
  error,'' \emph{IEEE Transactions on Information Theory}, vol.~46, no.~3, pp.
  1068--1077, May 2000.

\bibitem{Gabidulin-OptimalArrayCorrectingCodes_1985}
E.~M. Gabidulin, ``Optimum codes correcting lattice errors,'' \emph{Problemy
  Peredachi Informatsii}, vol.~21, no.~2, pp. 103--108, 1985.

\bibitem{Roth-ProbabilisticCrisscrossErrorCorrection_1997}
R.~M. Roth, ``{Probabilistic crisscross error correction},'' \emph{IEEE
  Transactions on Information Theory}, vol.~43, no.~5, pp. 1425--1438, Sep.
  1997.

\bibitem{wachterzeh2017listdecodingcrisscross}
A.~{Wachter-Zeh}, ``List decoding of crisscross errors,'' \emph{IEEE
  Transactions on Information Theory}, vol.~63, no.~1, pp. 142--149, 2017.

\bibitem{bitar2020criss}
R.~Bitar, I.~Smagloy, L.~Welter, A.~Wachter-Zeh, and E.~Yaakobi, ``Criss-cross
  deletion correcting codes,'' \emph{arXiv preprint arXiv:2004.14740}, 2020.

\bibitem{ISITAcriss}
------, ``Criss-cross deletion correcting codes,'' \emph{International
  Symposium on Information Theory and Its Applications (ISITA)}, 2020.

\bibitem{manabu2020delarrays}
M.~Hagiwara, ``Conversion method from erasure codes to multi-deletion
  error-correcting codes for information in array design,'' \emph{International
  Symposium on Information Theory and Its Applications (ISITA)}, 2020.

\bibitem{Kulkarni_Nonasymptotic-Upper-Bounds-for-Deletion-Correcting-Codes_13}
A.~A. {Kulkarni} and N.~{Kiyavash}, ``Nonasymptotic upper bounds for deletion
  correcting codes,'' \emph{IEEE Transactions on Information Theory}, vol.~59,
  no.~8, pp. 5115--5130, Aug 2013.

\bibitem{Gabidulin_TheoryOfCodes_1985}
E.~M. Gabidulin, ``Theory of codes with maximum rank distance,'' \emph{Problemy
  Peredachi Informatsii}, vol.~21, no.~1, pp. 3--16, 1985.

\end{thebibliography}
